\documentclass[journal]{IEEEtran}
\usepackage{cite}
\usepackage{amsmath,amssymb,amsfonts}
\usepackage{algorithmic}
\usepackage{graphicx}
\usepackage{textcomp}
\usepackage{xcolor}
\usepackage{algorithm,algorithmic}

\usepackage{subcaption} 
\usepackage{bbding}
\usepackage{pifont}
\usepackage{wasysym}
\usepackage[useregional]{datetime2}
\usepackage{fancyhdr}
\usepackage{hyperref}
\usepackage{cancel}
\usepackage{setspace,lipsum}

\ifCLASSINFOpdf
\else
\fi

\begin{document}
%
\title{A Multi-Agent DRL-Based Framework for Optimal Resource Allocation and Twin Migration in the Multi-Tier Vehicular Metaverse}
%
%
%

\author{Hayla~Nahom~Abishu,~\textit{Member, IEEE},~A.~Mohammed~Seid, ~\textit{Member, IEEE}, ~Aiman~Erbad,~\textit{Senior~Member,~IEEE}, ~Tilahun~M.~Getu,~\textit{Member, IEEE},~ Ala~Al-Fuqaha, ~\textit{Senior~Member, IEEE},~and ~Mohsen~Guizani,~\textit{Fellow, IEEE}
\thanks{H. N. Abishu, A. M. Seid, and A. Al-Fuqaha are with the Division of Information and Computing Technology,
College of Science and Engineering, Hamad Bin Khalifa University, Doha, Qatar.}
\thanks{A. Erbad is with the College of Engineering, Qatar University, Doha, Qatar.}
\thanks{\IEEEcompsocthanksitem T. M. Getu is with the Electrical Engineering Department, \'Ecole de Technologie Sup\'erieure (\'ETS), Montr\'eal, QC H3C 1K3, Canada (e-mail: tilahun-melkamu.getu.1@ ens.etsmtl.ca).}
\thanks{M. Guizani is with Machine Learning Department, Mohamed Bin Zayed University of Artificial Intelligence (MBZUAI), Abu Dhabi, UAE. }
}

\markboth{}%
{Shell \MakeLowercase{\textit{Hayla et al.}}: Vehicular
Twin Migration in the Multi-Tier Vehicular
Metaverse}

\maketitle

\begin{abstract}
 
Although multi-tier vehicular Metaverse promises to transform vehicles into essential nodes -- within an interconnected digital ecosystem -- using efficient resource allocation and seamless vehicular twin (VT) migration, this can hardly be achieved by the existing techniques operating in a highly dynamic vehicular environment, since they can hardly balance multi-objective optimization problems such as latency reduction, resource utilization, and user experience (UX). To address these challenges, we introduce a novel multi-tier resource allocation and VT migration framework that integrates Graph Convolutional Networks (GCNs), a hierarchical Stackelberg game-based incentive mechanism, and Multi-Agent Deep Reinforcement Learning (MADRL). The GCN-based model captures both spatial and temporal dependencies within the vehicular network; the Stackelberg game-based incentive mechanism fosters cooperation between vehicles and infrastructure; and the MADRL algorithm jointly optimizes resource allocation and VT migration in real time.  By modeling this dynamic and multi-tier vehicular Metaverse as a Markov Decision Process (MDP), we develop a MADRL-based algorithm dubbed the Multi-Objective Multi-Agent Deep Deterministic Policy Gradient (MO-MADDPG), which can effectively balances the various conflicting objectives. Extensive simulations validate the effectiveness of this algorithm that is demonstrated to enhance scalability, reliability, and efficiency while considerably improving latency, resource utilization, migration cost, and overall UX by 12.8\%, 9.7\%, 14.2\%, and 16.1\%, respectively.

\end{abstract}


\IEEEpeerreviewmaketitle

\section{Introduction}
\subsection{Motivation}
The Metaverse, in the realm of current and next-generation networks, is expected to revolutionize how we interact with digital environments.  These networks will offer ultra-fast, low-latency, and dependable connectivity, crucial for immersive experiences\cite{9880528,lee2024all,10373900}. The Metaverse has been integrated into vehicular networks to provide immersive and interactive vehicular services, such as onboard information and entertainment systems where people may interact, learn, collaborate, and play \cite{10376398}. As the world becomes increasingly connected, vehicles are evolving into integral components of a vast digital environment, transcending their traditional roles as mere modes of transportation. This evolution is being driven by advances in communication technologies, artificial intelligence, and edge computing, leading to the development of a dynamic and interconnected network known as the \textit{vehicular Metaverse}\cite{10273374}.
 Vehicular Metaverse has the potential to transform various aspects of the automotive industry, such as enhanced driving experiences through Augmented Reality (AR), Virtual Reality (VR), and Mixed Reality (MR), which can provide drivers with real-time information about their surroundings, including traffic updates, navigation assistance, and points of interest \cite{anwar2024moving}. The vehicular Metaverse integrates advanced technologies into the automotive industry; improves the driving experience; provides new forms of entertainment; and revolutionizes the way we interact with our vehicles. In vehicular Metaverses, each Vehicular Metaverse User (VMU) seamlessly connects with their Vehicular Twin (VT) to unlock immersive experiences. The VMUs continuously gather real-time data such as current location, historical trajectories, and service preferences\cite{10269659,10537999}. This information is used to generate VT tasks, ensuring seamless VT synchronization and enabling precise as well as dynamic interactions between the physical and virtual realms.
 
Because the vehicles have limited local computation and storage resources to execute computation-intensive VT tasks\cite{10019579}, they migrate VTs into nearby Road Side Unit (RSU) to perform large-scale VT execution and synchronization tasks\cite{10505943}. VT migration involves transferring a vehicle's digital twin virtual representation across different computational nodes in the Metaverse architecture. This process ensures that the digital twin remains synchronized with the physical vehicle, while maintaining high performance, despite changes in network conditions\cite{10396839}. 
However, due to vehicle mobility and limited coverage of RSUs, vehicles could be far from their VTs, increasing communication latency and reducing the Quality of Experience (QoE) of VMUs. Furthermore, the exponential growth in the number of vehicles and the massive data they generate places a significant strain on  RSUs, which have limited computational and communication resources\cite{9832009}. This means the resources required to execute their VT tasks may exceed the capacity of the source RSU servers in the coverage area. Besides, the source servers may malfunction or become overburdened to deliver the services required by VMUs.  

Concerning the mentioned scenario, recent studies have explored collaborative strategies among multiple RSUs to enhance resource allocation and VT migration in vehicular Metaverses \cite{10183802}. For instance, a coalition game-based approach has been proposed to ensure reliable VT migration, where RSUs form coalitions to jointly provide bandwidth resources, improving the QoE for VMUs\cite{10281020}. Meanwhile, the study in \cite{10505943} introduces a Multi-Agent Deep Reinforcement Learning (MADRL)-based Stackelberg game model, incorporating social-awareness and queueing theory to optimize VT migration. 
Furthermore, a multi-attribute auction-based resource allocation mechanism has been proposed to optimize resource allocation during VT migration considering both price and nonmonetary attributes, such as location and reputation, to enhance the efficiency of VT migration in vehicular Metaverses\cite{10608164}. These collaborative strategies among RSUs, utilizing game-theoretic approaches, are essential for optimizing performance and enhancing User Experience (UX) in vehicular Metaverses. 

The existing VT migration and resource allocation solutions face significant challenges in effectively coordinating resources across vehicles, edge, and cloud layers due to dynamic mobility, fluctuating network conditions, and complex multi-tier architecture\cite{10746337}. Those approaches overlook hierarchical integration and fail to fully utilize resources across these tiers, resulting in performance bottlenecks\cite{10700699}. More specifically, cloud resources remain underutilized due to latency challenges and the lack of effective resource allocation and VT migration strategies that adequately integrate vehicle-specific factors such as location, speed, and connectivity into decision-making processes. This may result in increased delays, migration costs, and inefficient resource utilization in multi-tier Metaverse environments\cite{10689465}.  Many of the existing resource allocation and VT migration solutions fail to address the need for a fair and efficient incentive scheme that motivates system entities to engage actively and truthfully\cite{10829636}. Additionally, the lack of real-time VT synchronization disrupts seamless connectivity between vehicles, RSUs, and virtual environments, making traditional approaches ineffective in addressing the stringent demands of the vehicular Metaverse.  
\subsection{Contributions}
We propose a novel multi-tier resource allocation and VT migration framework that integrates Graph Convolutional Networks (GCNs), a hierarchical Stackelberg game-based incentive mechanism, and MADRL to enhance the Metaverse service provisioning.  
We utilize GCNs in the proposed solution to model spatial and temporal dependencies in multi-tier vehicular Metaverses, capturing interactions among vehicles, edge, and cloud servers. This GCN-based vehicle-centric solution considers vehicle-specific attributes like speed, location, and connectivity to prioritize real-time needs in resource allocation and VT migration. By analyzing spatial dependencies and communication patterns in multi-tier vehicular Metaverses, the proposed solution predicts resource demands and identifies optimal migration paths for VTs. This approach models the vehicular network as a graph, where the nodes represent vehicles, edge servers, and cloud servers, and the edges capture communication links. Our proposed approach dynamically learns and optimizes decisions to ensure seamless VT services and efficient utilization of resources. In this hierarchical structure, each tier allocates resources based on real-time demands while optimizing for multiple objectives, such as migration success rate, UX, latency, and resource utilization. In addition, a game-theoretic incentive mechanism encourages cooperation among vehicles and network entities, aligning their objectives toward improving the overall system performance.
To solve the joint resource allocation and VT migration optimization problem, we proposed a MADRL-based approach, called Multi-Objective Multi-Agent Deep Deterministic Policy Gradient (MO-MADDPG) algorithm, which works in tandem with the GCN to optimize resource allocation and twin migration decisions. The MO-MADDPG framework leverages multi-agent reinforcement learning to handle the complexity of multi-tier vehicular networks, enabling agents to learn and adapt to the constantly changing environment. With the GCN providing insights into the network structure and dynamic interactions, the agents can make informed decisions regarding resource allocation and migration paths. The main contributions of this work are:
\begin{itemize}
    \item We propose a dynamic VT migration and resource allocation framework for the multi-tier vehicular Metaverse, integrating GCNs, a hierarchical Stackelberg game, and MADRL to enable real-time decision-making and efficient resource management by analyzing traffic patterns, road conditions, resources, and vehicle characteristics. 
\item We introduce a hierarchical Stackelberg game-based incentive mechanism for strategic decision-making, where entities across multiple layers (vehicle, edge, and cloud layers) set their strategies and respond optimally in the system. 
\item We formulate a joint VT migration and resource allocation optimization problem as a Markov Decision Process (MDP) and adopt a MO-MADDPG algorithm to solve the formulated optimization problem. 
\item We evaluate the performance and scalability of the proposed approach using simulations and real-world experiments.
\end{itemize}
The remainder of this paper is organized as follows. Section II reviews related works. Section III describes the system model. Section IV formulates the optimization problem; presents the proposed MADRL-based solution. Section V discusses the performance evaluation. Section VI concludes this work.
\section{Related works}
 
\subsection{Resource Allocation in Metaverse System}
The work in \cite{10250875} investigated resource allocation strategies within an AR-enabled vehicular edge Metaverse, focusing on maximizing the Metaverse operator's reward by jointly optimizing the CPU frequency and transmit power of AR vehicles, the sizes of computation models, and the distribution of computational resources on the Metaverse server.
  The investigation in \cite{10144631} proposed an adaptive edge resource allocation method based on Soft Actor-Critic with GCN (SAC-GCN), which defines the multiuser Metaverse environment as a graph with each agent represented by a node. The study in \cite{9880566} introduced a hierarchical game-theoretic-based coded distributed computing (CDC) framework to provide collaborative computing and Metaverse services for the vehicular Metaverse, where idle resources of vehicles can be allocated to handle intensive computation tasks. In the same study, the authors develop a coalition game to select reliable and resource-rich vehicles based on the reputation values calculated through the subjective logical model; a Stackelberg-based incentive mechanism to motivate a coalition of vehicles to participate in rendering tasks. 
  The authors of \cite{10368052} presented Human Centric (UC) resource allocation, considering joint communication and computational resources as well as VR video resolutions. The authors addressed the non-convex system cost problem, which includes Energy Consumption (EC) and delay, and solved it using the fractional programming technique. The work in \cite{10148094} proposed a Joint Resource Allocation and Metaverse Service (JRAMS) selection strategy that dynamically allocates communication and computation resources to Metaverse services with high QoE. This study proposed \textit{meta-distance} -- i.e., a novel metric for measuring virtual distance in the Metaverse that considers both service latency and social distance among users. JRAMS consists of two steps: i) a one-to-many matching game with externalities to match base stations and Metaverse users using Non-Orthogonal Multiple Access (NOMA) subchannel allocations, and ii) a coalition formation game to solve the Metaverse service selection problem. 
 \subsection{Vehicular Twin Migration}
In addressing Metaverse service interruption due to the movement of vehicles and the limited-service coverage of RSUs, some studies present VT migration solutions, where RSUs contribute resources to host VTs and facilitate efficient VT migration. Since a single RSU is insufficient for supporting large-scale VTs migration. The authors of \cite{10281020} proposed a coalition game approach framework for reliable VT migration in vehicular Metaverses. The coalition game among RSUs is formulated based on the reputation value of RSUs computed through the subjective logic model. 
The RSUs in the coalition game collectively provide bandwidth resources for reliable and large-scale VTs migration, while serving several VTs migrations at the same time.
 The authors of \cite{zhong2023blockchain} presented a blockchain-assisted game approach for VT migration in vehicular Metaverse to maintain continuous service and offer immersive experiences for VMUs as vehicles move. In this framework, the reputation value of RSUs is calculated based on their interaction freshness with VMUs, and the coalition of RSUs is formed based on their reputation value to share bandwidth resources. To provide immersive Metaverse services and efficiently migrate the VTs,  the coalition with the maximum utility is chosen, and a Stackelberg model is adopted to handle the interaction between the RSUs' coalition and VMUs, motivating VMU's active participation. This aims to provide seamless VT migration and enhance Metaverse services for VMUs. 
 
 To address the problems with the VT migration process due to insufficient bandwidth resources from RSUs for timely migration, the study in \cite{10505943} presented a Stackelberg-based incentive mechanism for vehicular Metaverse and proposed a MADRL algorithm, Multi-Agent LSTM-based Proximal Policy Optimization (MALPPO). The RSU that provides the Metaverse service and the RSU that requests the Metaverse interact through this incentive game.  The MALPPO algorithm facilitates learning the \textit{Stackelberg Equilibrium (SE)} without requiring private information from others, relying solely on past experiences. The work in\cite{10415630} presented an avatar task migration approach based on MADRL to address the limited resources available on vehicles and the high mobility of vehicles, where the avatar task is migrated to the nearest RSU or UAV for execution, leading to reduced communication overhead and task processing latency.  The authors utilized the MAPPO algorithm to optimize the avatar migration optimization problem. 
 
 To improve the convergence of MAPPO due to dimensionality and nonstationary problems in sharing parameters, the work in \cite{10415630} applied a Transformer-based MAPPO approach via sequential decision-making models, and a smart contract with blockchain is utilized to ensure trustworthiness in the computation resource-sharing transactions.  To address the challenges of making avatar migration decisions due to vehicle mobility, dynamic workload of RSUs, and RSU heterogeneity, the authors in \cite{10185562} proposed a MADRL-based dynamic avatar task migration framework based on real-time trajectory prediction. Specifically, they proposed a model to predict the future trajectories of vehicles based on their historical data, which could be useful for forecasting the available resources and workload of RSUs. The avatar task migration problem is formulated as a long-term mixed-integer programming problem, which is then transformed into a partially observable MDP, and multiple DRL agents with hybrid continuous and discrete actions are used to address the formulated optimization problem.

\begin{table*}[http]
	\scriptsize
    \caption{The comparison of our work with state-of-the-art works}
	\label{tab.1}
	\centering
	 \begin{tabular}{|p{0.8cm}| p{3.9cm}| p{3.6cm}| p{3.5cm}| p{3.8cm}| p{4.5cm}| }
    \hline
  \textbf{Paper} & \textbf{Scenario}&\textbf{Method}&\textbf{Problem}&\textbf{Objective}\\
		\hline
   \cite{9973630} & Blockchain-based mobile edge computing platform for resource sharing.&RL algorithm for multiple task allocation. & Multiple task allocation.
 & Optimize costs and utility. \\
   \hline
   \cite{10144631} & Adaptive edge resource allocation in the Metaverse using SAC-GCN.&Multiagent SAC-GCN with self-attention mechanism. & Edge resource allocation for multiple MVU. & Optimizing resource allocation and utilization rate. Improves UX. 
\\
		\hline
  \cite{9880566} &Collaborative computing paradigm based on CDC in vehicular Metaverse. & Game-theoretic CDC. Coalition formation game and Stackelberg game.  &Real-time rendering inefficiency
and reputation management.
 &Improving UX and resource utilization rate. \\
  \hline
\cite{10368052} & Human-centric resource allocation in the Metaverse over wireless communications. & Fractional programming technique for non-convex optimization. UC utility measure. &Insufficient bandwidth, power, computing, resolution, and CPU frequency resources.
 &Optimizing resource allocation and maximizing utility-cost ratio.\\
\hline
 \cite{10148094} & QoE analysis and resource allocation for wireless Metaverse services. &Personalized resource and attention-aware rendering capacity allocation. & Metaverse
service selection and resource allocation. & Enhancing QoE in Metaverse services.
\\
		\hline

  \cite{10281020} & Coalition formation game for VT migration in vehicular edge computing. & Double-level coalition formation game. & Limited resource of single RSU for VT migration.& Minimizing expenses and maximizing revenue.\\
		\hline
  \cite{zhong2023blockchain} & Blockchain-assisted twin migration for vehicular Metaverses.
 &MADRL algorithm for migration decisions. & Avatar task migration in vehicular Metaverses for immersive services.& Optimize avatar task migration by integrating learning-based algorithms.
\\
		\hline
  \cite{10505943} & VT migration in vehicular Metaverse. &MADRL with Stackelberg game. & VT migration.&  Optimize resource allocation.\\
		\hline
 \cite{10415630} & UAV-assisted avatar task migration for vehicular Metaverse. &MAPPO algorithm for the avatar task migration.& Avatar task migration from vehicles to RSUs/UAVs. & 
Latency reduction and UX. 
\\
		\hline
  \cite{10185562} & Avatar migration in vehicular Metaverses.& MADRL  for real-time trajectory prediction and avatar task migration. & Inefficient avatar migration decisions. & Minimize avatar service latency with an optimized pre-migration decision. 
\\
		\hline
Our work & Resource allocation and VT migration in the multi-tier vehicular Metaverse.&MO-MADRL and hierarchical Stackelberg game-based incentive. & Inefficient resource allocation and VT migration decisions.  & Minimizing migration costs, EC, and latency while maximizing UX.
\\
		\hline
	\end{tabular}
\end{table*}

We present a comparison and summary of the most relevant existing works with our proposed framework in TABLE \ref{tab.1}. As tabulated, the studies in \cite{9973630,10144631,9880566,10368052,10148094} focused on allocating computation and communication resources to enhance the immersive Metaverse experience and resource utilization efficiency in a variety of Metaverse services application scenarios. Furthermore, the works in \cite{10281020,zhong2023blockchain,10505943,10415630,10185562} investigated avatar tasks and VT migration in a variety of application contexts to provide reliable and efficient Metaverse services. However, the mentioned works considered neither a multi-tier resource allocation nor a VT migration in vehicular Metaverse. Aiming to close this knowledge gap and advance the field meaningfully, we propose a joint resource allocation and VT migration framework that can dynamically allocate resources and relocate the VTs across different layers of the multi-tier vehicular Metaverse so as to optimize performance, reliability, and resource utilization. The proposed framework can dynamically allocate resources and relocate the VTs across different layers of the multi-tier vehicular network to optimize performance, reliability, and resource utilization. \newline 
We now proceed to detail our system model.

\section{System Model}
\begin{figure*}[!t]
\centerline{\includegraphics[width=5.5in]{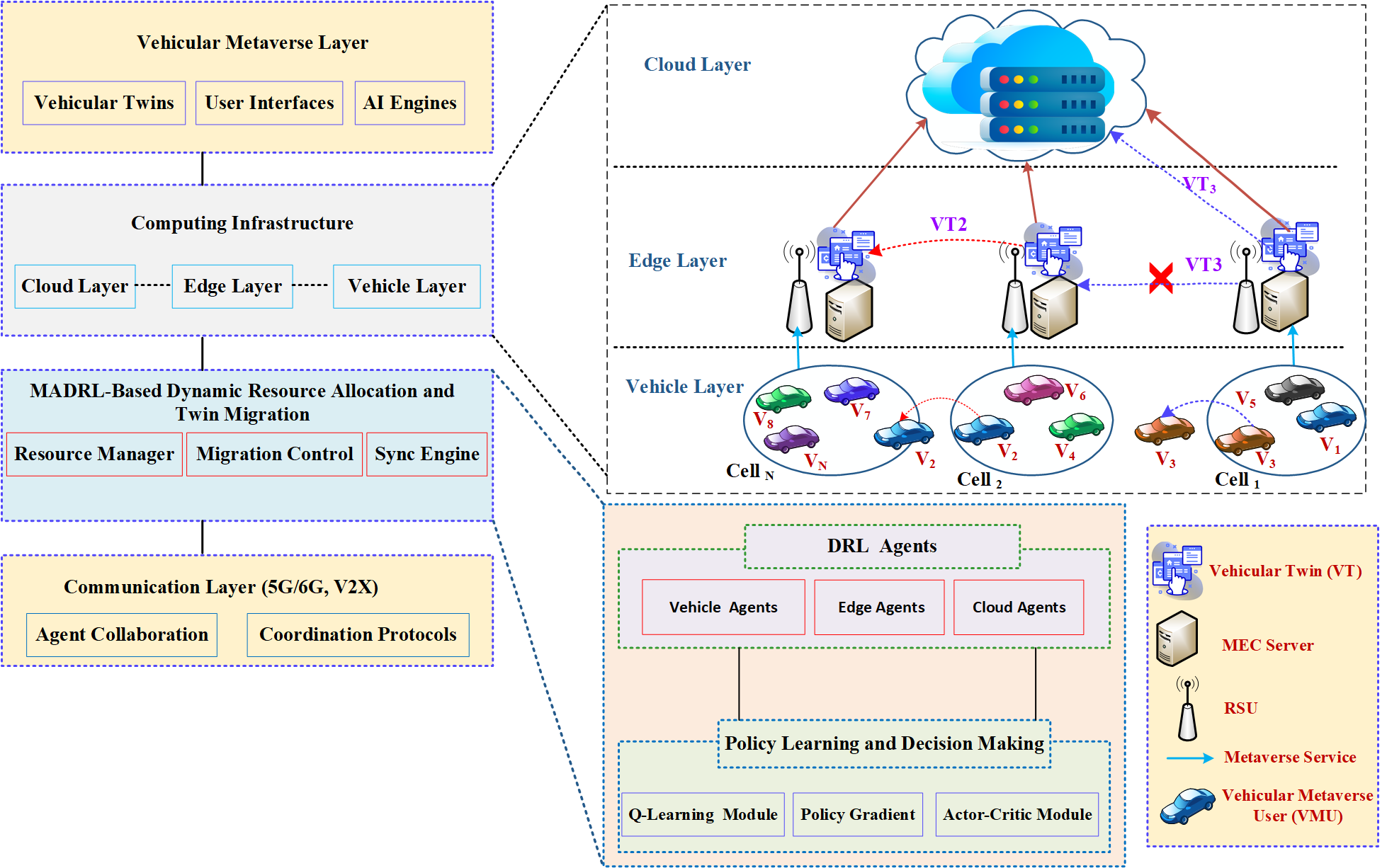}}
\caption{System model for a vehicle-centric resource allocation and a VT migration.}
    \label{fig:sys}
\end{figure*}

Our system model is shown in Fig. \ref{fig:sys} that schematizes a joint resource allocation and VT migration framework across a three-tier network -- comprising vehicle, edge, and cloud layers -- providing real-time data processing, analysis, and immersive Metaverse services. 

We consider a vehicular Metaverse system with different Metaverse services, such as healthcare services, augmented navigation, and immersive infotainment services. We define the sets of source edge nodes (i.e., RSUs), destination edge/cloud servers, and vehicles owned by VMUs as $\mathcal{J}:=\{1, \dots, J\}$, $\mathcal{I}:=\{1, \dots, I\}$, and $\mathcal{V}:= \{1, \dots, V\}$, respectively. In this scenario, Metaverse Service Providers (MSPs) own and manage the resources of both the source and destination servers. The set of MSPs is defined as $\mathcal{P} := \{1, \dots, P\}$. Each MSP provides one type of Metaverse service to VMUs with advanced AR/VR technologies, resulting in high-quality and immersive experiences, and also provides Metaverse resources when needed. 
Each vehicle is equipped with sensors, communication modules, and computational capabilities and has a VT representing its state, which includes sensor data, operational status, and contextual information. 

We consider the VTs require computational resources to process data, run applications, and communicate with other components in the Metaverse, whenever it is challenging to execute these services with local resources. The edge layers consist of RSUs located close to the vehicles to provide low-latency computation and storage. Still, edge layer resources are limited, and RSU coverage can be short,  making it difficult to serve moving vehicles. When vehicles leave the coverage area of the source server or the source server lacks sufficient resources to complete the VT tasks, these tasks can be migrated to another server on the nearest RSU or cloud servers. In this process, the source server, the vehicle, and the destination servers collaborate via a hierarchical Stackelberg game-based incentive platform \cite{10082939,10783015}, enabling optimal resource allocation and an efficient VT migration strategy. In this regard, cloud data centers are considered to provide reliable resources and Metaverse services with seamless VT migration across multiple layers. The cloud servers are equipped with a large amount of communication, computing, and storage resources to execute and store computation-sensitive VT tasks in the vehicular Metaverse system. This can relieve resource constraints on the edge layer while maintaining reliable Metaverse services.
\subsection{GCN-based Multi-Layer Network Modeling}
In our scenario, GCN is employed to model the multi-tier vehicular network as a graph \( G := (V, E) \), where \( V \) represents nodes across three layers of the hierarchy: vehicle, edge, and cloud. Also, \( E \) represents the edges that define the communication links between these layers, including vehicle-to-edge and edge-to-cloud interactions. The GCN extracts topological features and vehicle mobility patterns by processing node features through layers, generating embeddings that inform resource allocation and VT migration decisions. Each node $v_x$ is characterized by a feature vector $\mathcal{F}_x$, containing parameters like CPU and bandwidth utilization, latency, migration cost, EC, and UX. The GCN processes these features using a layer-wise propagation rule, where the updated feature matrix at the ($l+1$)-th layer is calculated as\cite{10066166}:
\begin{equation}\label{eq-gcn}
   H^{l+1}:=\sigma\bigg(\Bar{D}^{-1/2}\Bar{A}\Bar{D}^{-1/2}H^{(l)}W^{(l)}\bigg),
\end{equation}
 where \( H^{(l)} \) represents the node feature matrix at layer \( l \), \( \Bar{A} = A + I \) is the adjacency matrix with added self-loops, \( \Bar{D} \) corresponds to the degree matrix of \( \Bar{A} \), \( W^{(l)} \) is the trainable weight matrix, and \( \sigma \) denotes an activation function, such as ReLU. In this expression, $I$ is an identity matrix of the same dimensions as the adjacency matrix $A$, ensuring self-loops that stabilize learning and preserve node features in GCN. The GCN generates node embeddings $H^{(L)}$ at the final layer: $H^{L}:=\sigma\bigg(\Bar{D}^{-1/2}\Bar{A}\Bar{D}^{-1/2}H^{(L-1)}W^{(L-1)}\bigg)$. These embeddings $H^{(L)}$ serve as inputs to a MADRL framework to solve the multi-objective optimization problem by incorporating dynamic metrics such as UX $\mathcal{E}$, latency $L$, EC $E$, utility $\mathcal{U}$, and migration cost $C$ into the node feature $\mathcal{F}_x$, where $x\in\{1,2,\dots,v,\dots,V,1,2,\dots,j,\dots,J,1,2,\dots,i,\dots,I\}$. Here, the node feature $\mathcal{F}_x$ is defined as $\mathcal{F}_x:=\big[\mathcal{E}_v,\mathcal{U}_x,L_{v,j,i},E_{v,j,i},C_{v,j,i}\big]$, where $\mathcal{E}_v$, $\mathcal{U}_x$, $L_{v,j,i}$, $E_{v,j,i}$, and $C_{v,j,i}$ denote the vehicle UX, total utility (comprising the utility of the vehicle $\mathcal{U}_v$, the utility of the edge node $\mathcal{U}_j$, and the utility of the cloud node $\mathcal{U}_i$), latency associated with VT migration, EC, and the migration cost, respectively. This GCN-based multi-tier model effectively captures the dynamics and interactions in the vehicular Metaverse, enabling better resource allocation and migration decisions \cite{10144631}. 
\subsection{Channel Allocation and Communication Model}
In our proposed multi-tier architecture, we consider Orthogonal Frequency Division Multiplexing (OFDM) for vehicular Metaverse communication\cite{HSCL05}. The vehicles can enjoy the immersive Metaverse services through their VTs. To achieve reliable Metaverse services, it is essential to ensure seamless migration and synchronization of VT tasks while optimizing resource allocation. Our proposed resource allocation and VT migration strategy consider two scenarios: 1) vehicle-to-edge or cloud server migration, and 2) edge server-to-cloud server migration. In the first scenario, the vehicles can migrate their VTs to edge servers on RSUs or cloud servers owned by MSPs. The edge server $j$ or the cloud server $i$ is selected based on its resource availability and reliability, ensuring the VT task remains uninterrupted. In the second situation, if the VT is migrated to the nearest edge server $j$ and the vehicle moves outside of its coverage area, the VT task must be relocated to another edge server or cloud server $i$ owned by MSP $p$. In this case, once the edge node $j$ receives the VT input task from the vehicle $v$, creates a  VT task profile $\Upsilon_v:=\{D_v,\Omega, L_{max}\}$ to migrate to the cloud servers or another edge node, where $D_v$ is the VT data including the real-time VT states, historical interaction data, and vehicle configuration, $\Omega$ is the required CPU cycles, and $ L_{max}$ is the maximum tolerable delay. Let $\chi_{v,j,i}\in\{0,1\}$ be a binary migration decision variable and $\chi_{v,j,i}\in\{\chi_{v,j,i}^{v\rightarrow j},\chi_{v,j,i}^{v\rightarrow i},\chi_{v,j,i}^{v\rightarrow j\rightarrow i}\}$. This can be expressed as:
\begin{align}
   \chi_{v,j,i}(t):=\begin{cases}
        \chi_{v,j}^{v\rightarrow j}(t)=1, \text{vehicle}\quad v \quad\text{migrates its}\quad \Upsilon_v \\ \text{to the nearest edge server} \quad j\\
        \chi_{v,j}^{v\rightarrow i}(t)=1, \text{vehicle}\quad v \quad\text{migrates its}\quad \Upsilon_v \\ \text{to cloud server} \quad i.\\
        \chi_{v,j,i}^{v\rightarrow j\rightarrow i}(t)=1, \text{source server}\quad j \quad\text{migrates}  \\ \Upsilon_v \quad \text{of vehicle} \quad v\quad \text{to cloud server} \quad i,
    \end{cases}
\end{align}
where $\chi_{v,j,i}^{v\rightarrow j}+\chi_{v,j,i}^{v\rightarrow i}+\chi_{v,j,i}^{v\rightarrow j\rightarrow i}=1$. This decision variable helps vehicles and resource-providing servers to make VT migration and resource allocation decisions. The edge nodes (RSUs) or cloud servers can allocate resources, i.e., bandwidth and computing resources based on agreed-upon monetary incentives employing MDRL-supported Stackelberg game-based decision-making strategies. This hierarchical decision-making ensures optimal resource utilization while maintaining service quality and meeting stringent requirements of VT task execution. Let $\mathbb{C}$, $\mathcal{B}$, $ \Upsilon_v$, and $ B_v $ be the number of communication channels, the bandwidth of each channel, the VT of vehicle $ v \in V $, and the bandwidth requirements of VT task $\Upsilon_v$, respectively. For efficient channel and bandwidth resource allocation and utilization, the following constraints must be satisfied, which indicates that each vehicle can only be associated with a single edge/cloud server at a time slot $t$. 
\begin{equation}
    \sum_{j\in J} \sum_{i \in I} \chi_{v,j,i} \leq 1, \quad \forall v \in \mathcal{V},j\in \mathcal{J}, i\in \mathcal{I}.
\end{equation} 
Similarly, the bandwidth resource allocation must satisfy the constraint presented in (\ref{bw}), which implies that the resource requirements (bandwidth $B_v$) of each VT must be less than the bandwidth of the channel allocated to it. 
\begin{equation}\label{bw}
    \sum_{v \in V} \chi_{v,j,i} \times B_v \leq \mathcal{B}.
\end{equation}

Once the required resources are allocated, the vehicles offload their VT task inputs to the servers on the nearest RSUs or cloud data centers. The transmission latency of vehicle $v$ depends on the transmission rate of the channel between vehicle $v$ and server $j$ of MSP $p$, which is expressed as: $R_{v,j}^{tr}:=b\log_2\Big (1+\frac{\delta k d_{v,j}^{-\varepsilon}}{N_0}\Big)$, where the parameters $\delta, d_{v,j}, k, \varepsilon$, and $N_0$ denote transmission power of vehicle $v$, the distance between vehicle $v$ and server $j$, channel power gain, path-loss coefficient, and the Power Spectral Density (PSD) of the Additive White Gaussian Noise (AWGN), respectively. Thus, the transmission latency of VT task input of vehicle $v$ at time $t$ is calculated as:
\begin{align}
    L_{v}^{tr}(t):=\begin{cases}
        \chi_{v,j}^{v\rightarrow j}\frac{D_v}{R_{v,j}^{tr}}(t)\\
        \chi_{v,j}^{v\rightarrow i}\frac{D_v}{R_{v,i}^{tr}}(t),
    \end{cases}
\end{align} where $D_v$ and $R_{v,j}^{tr}$ are the VT task input size and the transmission rate between vehicle $v$ and edge server $j$, respectively, and $R_{v,i}^{tr}$ is the transmission rate between vehicle $v$ and cloud server $i$. To eliminate redundancy, the vehicle-to-cloud node interaction is excluded, as it follows a formulation similar to the previously mentioned equation. The VT tasks received from vehicles are either stored and executed by the source edge node or migrated to other edge nodes / cloud servers. 

\subsection{Computation Model}
In this multi-tier infrastructure, dynamically relocating VTs across the multi-tier network can optimize performance and resource utilization. This work aims to minimize the latency for processing VT tasks, balance the load across edge and cloud layers, and ensure real-time responsiveness and reliability for vehicular applications. Once the VT task is offloaded to the server $j$ and the server decides to execute it with its local resources, the execution latency of the VT task can be calculated as $L_j^{ex}(t):=\chi_{v,j}^{v\rightarrow j}\frac{\Omega D_v}{f_j}$, where $f_j$ is the allocated computational resource of the server $j$ of MSP $p$. 
In another case, when the source edge node faces resource limitations or detects that a vehicle has moved out of its coverage area, it migrates the VT tasks to a destination edge node or cloud server via the physical link between the source and destination nodes.  In this case, we consider transmission, queue, and re-instantiation delays.  
Thus, the transmission latency can be calculated as:
\begin{equation}
    L^{mig}_{j,i}(t):=\chi_{j,i}^{v\rightarrow j\rightarrow i}(t)\frac{D_v(t)}{R_{j,i}(t)},\forall v\in \mathcal{V}, j\in \mathcal{J},i\in \mathcal{I},
\end{equation}
where $R_{j,i}$ is the data transmission rate of the link between the source and destination nodes, which can be calculated as:
\begin{equation}
   R_{j,i}(t):=b_{j,i}\log_2(1+\frac{\Bar{\delta} \kappa d_{j,i}^{-\Bar{\varepsilon}}}{N_0}),
\end{equation} 
where $b_{j,i}$ is the bandwidth allocated for the channel between the source and distention nodes. The parameters $\Bar{\delta}, d_{j,i}, \kappa, \Bar{\varepsilon}$, and $N_0$ represent transmission power of source edge node $j$ of MSP $p$, the distance between source node $j$ and the $i$-th destination edge/cloud server, channel power gain, path-loss coefficient, and the PSD of the AWGN, respectively.
 After transmitting the VT task into the destination node, it might wait in a queue before it can be processed. The waiting delay can be affected by the network traffic and the processing capacity of the destination node. We consider an M/M/s with a non-preemptive priority queueing model to provide priority-based VT migration, where higher-priority VT tasks are served before lower-priority tasks, but once a task is being served, it cannot be interrupted. In this queuing model, the queuing delay calculation considers the arrival rates $\lambda$, service rates $\mu$, number of servers $s$, the number of VT tasks in queue $\Lambda_q$, and the prioritization of different VT tasks $\rho$. Thus, we calculate the queuing delay as: $L^{qu}_{i}(t):=\rho\frac{\Lambda_q}{\mu(s\mu-\lambda)}$. To restart the VT task processing on the new resource, the environment needs to be initialized and the VT task state needs to be loaded. Likewise, we calculate the re-instantiation delay as: $L_{i}^{rd}(t):=\frac{\Omega(t)}{f_{i}(t)}$, where $f_{i}$ is the computation capacity of the destination node or cloud server $i$. The overall VT migration delay of task $\Upsilon_v$ is given as: $L_{v,j,i}:= L_{v}^{tr}(t)+L_j^{ex}(t)+ L^{mig}_{j,i}(t)+L_{i}^{qu}(t)+L_{i}^{rd}(t)$. This objective function aims to minimize the total latency, which is expressed as:
\begin{align}\label{latency}
L:=\min \sum_{v\in V}\sum_{j\in J} \sum_{i\in I} \chi_{v,j,i} L_{v,j,i} \quad \qquad \qquad \quad\qquad \quad \quad\\
\textnormal{s.t.}\quad C1:\hspace{2mm}\sum_{v \in V} \chi_{v,j,i} \Omega_{\Upsilon_v} \leq \Omega_j,\Omega_{i}, \hspace{2mm}\forall v \in \mathcal{V},j\in \mathcal{J},i\in\mathcal{I} \nonumber\\
 C2:\hspace{2mm}\sum_{j\in J}\sum_{i\in I} \chi_{v.j,i} = 1, \hspace{2mm} \forall v\in \mathcal{V}, j\in \mathcal{J},i\in\mathcal{I} \qquad \nonumber\\
  C3:\hspace{2mm}L_{v,j,i} \leq L_{\text{max}},\hspace{2mm}\forall v \in \mathcal{V},j\in \mathcal{J},i\in\mathcal{I}, \qquad \quad \nonumber
\end{align}
In (\ref{latency}), the constraint $C1$ states that the CPU cycles required to process the VT task must be less than the CPU cycles of the source server $j$ and the destination node or the cloud server $i$; the constraint $C2$ enforces that the VT task can only be migrated to one destination edge/cloud server $i$; and the constraint $C3$ ensures that the total latency can not exceed the maximum latency threshold.
 We also evaluate the EC costs associated with the VT migration and update. We consider the VT task transmission energy $E_{v}$ required to transfer the task from the vehicle to the source edge node $j$ or cloud server $i$, the VT task execution energy $E_{j}^{ex}$ at the source edge node $j$, the VT task migration energy $E_{j,i}$ for transferring the task from the source node $j$ to destination edge/cloud server $i$, and the VT task execution energy $E_i$ at the destination edge/cloud server $i$. The EC during the migration of the VT task from vehicle $v$ to edge server $j$ or cloud server $i$ is given as:
\begin{align}
    E_v:= \begin{cases}
    \sum_{j\in J}\chi_{v,j}^{v \rightarrow j}\frac{P_vD_v}{R_{v,j}(t)} & \\
    \sum_{i \in I}\chi_{v,i}^{v\rightarrow i}\frac{P_vD_v}{R_{v,i}(t)}, 
    \end{cases}
    \end{align} 
    where $P_v$ is the transmission power of vehicle $v$.
The EC when executing the VT task at the edge server $j$ is calculated as $E^{ex}_{j}(t):=\sum_{v \in V}\chi_{v,j,i}\kappa_j\big(f_j(t)\big)^2$, where $k_j$ denotes CPU switch capacitance of the server $j$. The transmission EC incurred when migrating the VT task from the source edge server $j$ to the destination edge/cloud server $i$ can be expressed as: $E^{j\rightarrow i}_{j,i}:=\sum_{i \in I}\chi_{j,i}^{v\rightarrow j\rightarrow i}\frac{P_jD_v}{R_{j,i}(t)}$. After the VT task is migrated to the destination edge/cloud server $i$, we calculate the EC for executing the VT task as: 
\begin{align}
    E^{ex}_{i}:=\begin{cases}
        \chi_{v,i}^{v\rightarrow i}\sum_{v \in V}\kappa_i(f_i(t))^2 & \\
        \chi_{j,i}^{v\rightarrow j\rightarrow i}\sum_{j\in J}\kappa_i(f_i(t))^2.
    \end{cases}
\end{align}
Then, we calculate the total power consumption for maintaining Metaverse services and seamless VT migration as:
$E_{v,j,i}:=E_v+E^{ex}_{j}+E^{j\rightarrow i}_{j,i}+E^{ex}_{i}$ and the objective function is to minimize the EC, which is expressed as: 
\begin{align}\label{ec}
   E:= \min\sum_{v\in V}\sum_{j\in J}\sum_{i\in I}\chi_{v,j,i}E_{v,j,i}\qquad\\
    \textnormal{s.t.} \quad E_{v,j,i}\leq E_{max},\nonumber \qquad
\end{align}
 where $E_{max}$ is the maximum EC threshold for a VT task migration.   
\subsection{Vehicular Twin Migration Model}
The VT migration model is designed to ensure seamless service delivery and efficient resource utilization in dynamic vehicular environments within the multi-tier vehicular Metaverse. It facilitates the migration of VTs across vehicle,  edge, and cloud layers to maintain service quality as vehicles move between coverage areas\cite{10836864}. The VT migration is triggered when vehicles exit the coverage area of a source node, or when resource limitations or dynamic user demands necessitate task redistribution. The migration decision-making process considers factors like resource availability, latency, migration cost, and QoS requirements. The agents operating at the vehicle, edge, and cloud layers hierarchically interact through a Stackelberg game-based incentive mechanism to negotiate resource supply and pricing, enabling efficient resource allocation and seamless VT migration. We evaluate resource utilization at each layer and analyze the migration cost of transferring VT tasks, focusing its impacts on performance metrics like latency and service quality both during the migration process and after its completion. Thus, the migration cost of VT task $ \Upsilon_v $ is given as $C_{v,j,i}:=\Omega \theta_{j,i}$, where $\theta_{j,i}$ is the price for the resource allocated by the destination edge/cloud server $i$. This objective function aims to minimize the migration cost, which is given as:
\begin{align}\label{cost}
   C:= \min \sum_{v \in V} \sum_{j \in J}\sum_{i \in I}\chi_{v,j,i} C_{v,j,i} \qquad \qquad \qquad \qquad\quad \\
\textnormal{s.t.} \quad \chi_{v,j,i} \in \{0,1\}, \quad \forall v \in \mathcal{V}, j \in \mathcal{J},i \in \mathcal{I}, \nonumber \\ 
\sum_{j \in J} \sum_{i \in I} \chi_{v,j,i} \leq 1, \quad \forall v \in \mathcal{V},j\in \mathcal{J}, i \in \mathcal{I}.\nonumber
\end{align}
By optimizing these functions and adhering to the constraints, the system aims to achieve efficient resource allocation and seamless VT migration within the multi-tier vehicular Metaverse. 
\subsection{User Experience Model}
In this scenario, we also consider UX, which is a measure of user satisfaction that is often inversely related to latency but can also be influenced by various factors such as service quality and reliability. The vehicle UX during VT task migration across different tiers of the network can be expressed as:
$\mathcal{E}_v:=\chi_{v,j,i}\times\frac{1}{N}\sum_{n=1}^N\aleph_n$, where $\aleph_n$ is the UX rating for VT task $n$, which can be modeled as function of latency $f(L)$, service quality $f(\mathbb{Q})$ and reliability $f(\Re)$. We use a weighted sum to calculate experience rating as: $\aleph_n:=w_lf(L)+w_qf(\mathbb{Q})+w_rf(\Re)$, where $w_l, w_q,$ and $w_r$ are the weight parameters assigned to latency, service quality, and reliability, respectively, reflecting their relative importance. In this case, $f(L)$ negatively impacts UX and should be a decreasing function of latency $L$, which can be expressed as $f(L):=a_le^{-b_lL}$, where $a_l$ and $b_l$ are parameters that control the shape of the exponential decrease. The functions $f(\mathbb{Q})$ and $f(\Re)$ positively impact UX and should be increasing functions of service quality $\mathbb{Q}$ and reliability $\Re$, which can be expressed as $f(\mathbb{Q}):=a_q(1-e^{-b_q\mathbb{Q}})$ and $f(\Re):=a_r(1-e^{-b_r\Re})$, respectively. The parameters $a_q, b_q, a_r$, and $b_r $ control the shape of the exponential increase in service quality and reliability in the user's experience rating.  The objective function aims to maximize the vehicle UX as given below.
\begin{align}\label{UE}
    \mathcal{E}(L,\mathbb{Q},\Re):=\max\sum_{v \in V} \sum_{j \in J}\sum_{i \in I}\chi_{v,j,i} \mathcal{E}_v  \qquad \qquad\\
\textnormal{s.t.} \quad \mathcal{E}_v\geq \mathcal{E}_{min};\hspace{2mm} L\leq L_{max};\hspace{2mm}\Re\geq \Re_{min};\hspace{2mm} \mathbb{Q}\geq \mathbb{Q}_{min}. \nonumber
\end{align}
In  (\ref{UE}), the first constraint enforces that the vehicle UX must meet a minimum threshold to ensure satisfaction, while the second indicates that the latency must be less than the maximum threshold. Similarly, the third and the fourth constraints ensure that the reliability and service quality must be greater than their minimum threshold values, respectively. 
\subsection{Utility Function}
In this work, we consider the system utility that combines the utilities of individual participating nodes at each layer. We design a hierarchical Stackelberg game\cite{7879922,8004153} to structure the interaction between the vehicle, edge, and cloud layers, adopting a leader-follower framework where the decisions made at each layer impact not only its performance but also that of the other layers. 
This hierarchical Stackelberg game-based incentive structure consists of two distinct stages: 
\par
1) In the first stage, the destination cloud/edge servers act as leaders and the source edge servers act as followers. Let $\beta:=\{\beta_j,\beta_v\}$ represent the unit of resource required by edge node $j$ and vehicle $v$ to migrate/host a VT. Similarly, let $\theta:=\{\theta_i,\theta_j\}$ represent the corresponding resource prices, where $\theta_i$ is the price set by the cloud server $i$ for the resource demand $\beta_j$ of the source node $j$, and $\theta_j$ is the price set by edge server $j$ for the resource demand $\beta_v$ of vehicle $v$.
 The first stage of the game is formulated as follows.

\emph{Leaders' Pricing (Destination Cloud/Edge servers):}
 \begin{equation} \label{stg21}
  \begin{split}
       \max_{\beta,\theta\geq0}\mathcal{U}_i (\beta_j,\theta_i,\beta_v) \quad\\ 
       \textnormal{s.t.} \hspace{0.5cm} \sum_{v=1}^V\beta_v+\sum_{j=1}^J\beta_j\leq\chi_i,
  \end{split}
 \end{equation} 

 \par \emph{Followers' Demand (Edge Servers):}
 \begin{equation} \label{stg22}
    \begin{split}
        \max_{\beta,\theta\geq0}\mathcal{U}_j(\beta_v,\theta_j) \\ 
    \textnormal{s.t.}  \hspace{0.5cm} \sum_{v=1}^V\beta_v\leq\chi_j; \hspace{2mm}\ell_{j,i}\geq\ell^{min}_{j,i}.
    \end{split}
\end{equation}
\par
2) In the second stage, edge/cloud servers act as leaders by setting the prices for resources and VT migration/hosting, while vehicles--as followers--decide their resource and VT migration/hosting demands in response to the prices established by the leaders. The second stage game is formulated as:

\emph{Leaders' Pricing (Edge/Cloud Servers):}
 \begin{equation} \label{stg11}
  \begin{split}
       \max_{\theta\geq0}\mathcal{U}_j (\beta,\theta):=\max_{\theta\geq0}\mathcal{U}_j (\beta_j,\beta_v,\theta_j,\theta_c) \quad\\ 
       \textnormal{s.t.} \hspace{0.5cm} \sum_{v=1}^V\beta_v\leq\chi_j,
  \end{split}
 \end{equation} 

 \par \emph{Followers' Demand (Vehicles):}
 \begin{equation} \label{stg12}
    \begin{split}
        \max_{\beta\geq0}\mathcal{U}_v (\beta,\theta)\quad \\ 
    \textnormal{s.t.} \quad  \ell_{j,i}\geq\ell^{min}_{j,i}.
    \end{split}
\end{equation}
The system utility is calculated as $\mathcal{U}:=\sum_{v=1}^V\mathcal{U}_v+\sum_{j=1}^J\mathcal{U}_j+\sum_{i=1}^I\mathcal{U}_i$.
For this hierarchical Stackelberg game, we applied a backward induction method \cite{10783015} to identify the optimal strategy for each participant, ensuring the system reaches a stable equilibrium. In this process,  each agent adjusts its decisions sequentially, predicting the optimal choices of the others, and equilibrium is attained when all agents' strategies are mutually optimal.
 \par \textbf{Definition 1}: Let $\beta^*\in\{\beta^*_j,\beta^*_v\}$ and $\theta^*\in\{\theta^*_i,\theta^*_j\}$ be the optimal resource demand and the corresponding optimal price\cite{9838425}, respectively. The point ($\beta^*,\theta^*$) is the SE if it satisfies the following conditions.
\begin{equation}
        \mathcal{U}_i(\beta^*,\theta^*)\geq  \mathcal{U}_i(\beta_j,\theta^*_i). \qquad \qquad \qquad \qquad \qquad
\end{equation}
\begin{equation}
      \mathcal{U}_j(\beta^*,\theta^*)\geq  \mathcal{U}_j(\beta^*_j,\theta_i) \hspace{2mm}\&\hspace{2mm} \mathcal{U}_j(\beta^*,\theta^*)\geq \mathcal{U}_j(\beta_v,\theta^*_j).
\end{equation}
\begin{equation}
        \mathcal{U}_v(\beta^*,\theta^*)\geq \mathcal{U}_v(\beta^*_v,\theta_j). \qquad \qquad \qquad \qquad \qquad
\end{equation}
Given the complexity of the optimization problem, determining the optimal decision policy is difficult. To address this, we employ a multi-objective MADRL approach, specifically MO-MADDPG, to solve it. To confirm the existence and uniqueness of the SE in the game, we calculate the best response strategy for each player and derive the derivatives of the associated equations (\ref{stg11}), (\ref{stg12}), (\ref{stg21}), and (\ref{stg22}), with respect to $\beta$ and $\theta$ as in [\citen{9852754},~Eqs.~16,~17].

Since the objective functions are conflicting, we apply the weighted sum approach to combine them into a single optimization function that balances the conflicting factors.
\begin{equation}
    \mathbb{O}(t):=\omega_1\mathcal{E}+\omega_2\mathcal{U}-(\omega_3L+\omega_4E+\omega_5C),
\end{equation}
where $\omega_1,\omega_2,\omega_3$, $\omega_4$, and $\omega_5$ are the weights for UX, utility, latency, EC, and migration costs, respectively
\section{Problem Formulation and Proposed Solution}
\subsection{Problem Formulation}
The problems in our multi-tier vehicular Metaverse scenario are multifaceted, involving multi-objective optimization problems and constraints. These problems demand a systematic approach to balance objectives such as latency, EC, migration cost, and UX. The system leverages multiple MADRL agents distributed across the vehicle, edge, and cloud layers to collaboratively manage VT task migration and execution. These agents dynamically adapt to varying conditions, including network fluctuations, vehicular mobility, and resource availability, to ensure seamless task migration and optimal resource utilization. To achieve these objectives, we formulate the joint optimization problem based on (\ref{latency}), (\ref{ec}),(\ref{cost}), and (\ref{UE}), as follows.
 \begin{equation}
  \pmb{P}: \quad \max_{L,E,C,\mathcal{E},\mathcal{U}}\mathbb{O}(t)\qquad\qquad\qquad\qquad \qquad\qquad \qquad\qquad 
  \\   
 \end{equation}
   \begin{align}
\text{s.t.} \hspace{2mm} C1:& \hspace{2mm}chi_{v,j,i}\in \{0,1\},\forall v\in \mathcal{V}, j\in \mathcal{J},i\in\mathcal{I} \notag\\
C2:&\hspace{2mm}\sum_{j=1}^J \sum_{i=1}^I\chi_{v,j,i}\leq 1,\hspace{2mm}\forall v\in \mathcal{V}, j\in \mathcal{J},i\in\mathcal{I} \notag\\
C3:& \hspace{2mm}\sum_{v \in V} \chi_{v,j,i} \Omega \leq \Omega_j,\Omega_{i}, \forall v \in \mathcal{V},j\in \mathcal{J},i\in\mathcal{I} \notag\\
C4:&\hspace{2mm}\sum_{v \in V} \chi_{v,j,i} B_v \leq \mathcal{B},\forall v \in \mathcal{V},j\in \mathcal{J},i\in\mathcal{I} \notag\\
C5:&\hspace{2mm}L_{v,j,i} \leq L_{max},\forall v \in \mathcal{V},j\in \mathcal{J},i\in\mathcal{I} \notag\\
C6:&\hspace{2mm}E_{v,j,i}\leq E_{max},\forall v \in \mathcal{V},j\in \mathcal{J},i\in\mathcal{I} \notag\\
C7:&\hspace{2mm}C_{v,j,i} \leq C_{max},\forall v \in \mathcal{V},j\in \mathcal{J},i\in\mathcal{I} \notag\\
C8: & \hspace{2mm}\chi_{v,j,i}\mathcal{E}_v\geq \mathcal{E}_{min},\forall v \in \mathcal{V},j\in \mathcal{J},i\in\mathcal{I} \notag\\
C9:&\hspace{2mm}\Re\geq \Re_{min}\notag\\
C10:&\hspace{2mm}\mathbb{Q}\geq \mathbb{Q}_{min},\notag
\end{align} 
where $C1$ signifies a binary decision variable for VT task migration; $C2$ ensures that each vehicle can only associate with a single server at RSU or cloud server at a time; $C3$ and $C4$ affirm that the computing and bandwidth resource requirements of the VT task must be less than the allocated resources of source edge server j and cloud server i at a time t; $C5, C6, C7$, and $C8$ ensure that the latency, EC, migration cost, and UX meet the required threshold values; and $C9$ and $C10$ enforce the reliability and quality of services be greater than the minimum threshold values. The formulated optimization problem can be characterized as a \textit{Mixed-Integer Linear Programming (MILP)} problem due to its complex interplay of multiple objectives, diverse constraints, and the need to optimize across a mix of discrete and continuous decision variables. As a result, solving this MILP problem with conventional techniques is quite challenging, if not impossible. Furthermore, 
balancing the trade-offs between latency, migration cost, EC, and UX in a multi-tier vehicular Metaverse can be quite complex and hence \textit{NP-hard}. Thus, by incorporating advanced MADRL framework-based multi-objective optimization techniques, we can address this optimization problem, providing an optimal VT migration strategy for the continuous operation of VTs in a Metaverse environment, as detailed below.

\subsection{The Proposed MADRL-based Solution}
\subsubsection{MADRL-Based Framework}To handle the dynamics and complexity of decision-making in the vehicle-centric resource allocation and VT migration across multiple layers of the multi-tier vehicular Metaverse, we proposed a MADRL-based framework. This framework allows multiple agents at vehicle, edge, and cloud layers to make optimal decisions to balance the trade-off between latency, migration cost, EC, and UXs. The proposed MADRL-based algorithm can effectively deal with the multi-objective optimization problem to balance the trade-off between these conflicting objectives.
 We define the MDP with the tuple $(\mathcal{S}, \mathcal{A}, \mathcal{X},\mathcal{P},\mathcal{R},\gamma)$, where
$\mathcal{S}$ is the set of states, $\mathcal{A}$ denotes the set of action spaces, $\mathcal{P}(s'|s,a)$ is state transition probability function, $\mathcal{R}$ is a reward function, $\gamma$ denotes a discount factor, and $\mathcal{X}$ stands for the set of all agents across the Metaverse network tier and is defined as $\mathcal{X} := \mathcal{V} \cup \mathcal{J} \cup \mathcal{I}$.\newline
\emph{\textbf{State space:}} For the MADRL model to effectively manage resource allocation and VT migration across the dynamic multi-tier network, the state space is separated into the global state and local states. The global state and local states of the vehicle, edge, and cloud layers in the proposed framework are synergistic and hierarchical, enabling efficient decision-making in the dynamic multi-tier vehicular Metaverse. The global state of the vehicular Metaverse environment acts as an overarching view of the system integrating data from all local states and is defined as $\mathcal{S}:=\{\Gamma,\mathbb{T},\chi,\mathbb{P}\}$, where $\Gamma:=\{\mathcal{B},f\}$ is the aggregated resource availability (bandwidth $\mathcal{B}$, CPU cycle $f$) across the tiers, $\mathbb{T}$ is traffic patterns such as the dynamic behaviors of network and computational resource demands generated by vehicles and their digital twins, $\chi$ is connectivity status, and $\mathbb{P}:=\{E,\mathcal{E},L,\mathcal{C}\}$ is system-wide performance metrics across the vehicle, edge, and cloud layers. This global state is crucial for optimizing high-level objectives like EC ($E$), migration costs ($\mathcal{C}$), latency ($L$), and UX ($\mathcal{E}$).
Local states capture layer-specific parameters that are necessary for localized decision-making. At the vehicle layer, the local state $\mathcal{S}_V:=\{\rho,v,\Gamma_v,\chi,L_{max},C_{max},\mathbb{Q}_{min},\mathcal{E}_{min},\varsigma\}$ includes individual vehicle mobility information (e.g., position $\rho$, and velocity $v$), real-time resource demands $\Gamma_v:=\{\beta_v,f_v\}$, and connectivity status with nearby edge nodes $\chi$. It also includes VT requirements like latency constraints $L_{max}$, migration cost $C_{max}$, service quality $\mathbb{Q}_{min}$, UX $\mathcal{E}_{min}$, and VT status $\varsigma$ (e.g., current hosting server and migration requirements). At the edge layer, the local state $\mathcal{S}_E:=\{\Gamma_j,C_{max},\varsigma,L_{max},\mathcal{M}\}$ reflects the available resources $\Gamma_j$ of each edge node, migration costs ($C_{max}$), pending tasks and active VT instances, latency ($L_{max}$), and current load relative to capacity, where $\Gamma_j:=\{\mathcal{B}_j,f_j\}$, i.e., CPU cycle $f_j$, and bandwidth $\mathcal{B}_j$. For the cloud layer, the local state captures aggregate resource capacities of cloud servers, current resource utilization, migration costs between the cloud and other tiers, historical task performance data, and insights from feedback loops with edge nodes and vehicles.
In this way, the framework balances the high-level overview required for global optimization with the detailed, layer-specific insights necessary for precise decision-making. By leveraging GCNs, the framework effectively captures both spatial and temporal dependencies within the vehicular network, enabling accurate predictions of resource demands and optimal migration paths.
\newline
\emph{\textbf{Action space:}} The action space of all agents is represented as $\mathcal{A}:=\{\mathcal{A}_V, \mathcal{A}_E,\mathcal{A}_C\}$, where $\mathcal{A}_V,\mathcal{A}_E$, $\mathcal{A}_C$ correspond to the local actions of agents at the vehicle, edge, and cloud layers, respectively. We further define the actions of vehicle layer agents as $\mathcal{A}_V:=\{a_m,a_o,a_q\}$, where $a_m$, $a_o$, and $a_q$ denote the VT migration initiation (including resource request, mobility reporting), task offloading, and QoS determination actions, respectively. At the edge layer, the set of agents' actions can be defined as $\mathcal{A}_E:=\{a_r,a_m,a_l\}$, which involve resource allocation ($a_r$), VT hosting/migration ($a_m$), and communication links management ($a_l$) actions. Similarly, the agents' actions at the cloud layer can be defined as $\mathcal{A}_C:=\{a_r,a_g,a_m\}$, where $a_r,a_g,$ and $a_m$ denote high-capacity resource allocation for VT tasks, global state analysis, and VT hosting/migration management actions, respectively. Each agent acts according to the strategies it designed to achieve the optimization objectives. This hierarchical and distributed action framework ensures efficient collaboration between agents across layers to optimize resource allocation and seamless VT migration.

\emph{\textbf{Reward function:}} This function balances multiple conflicting objectives, including minimizing latency, EC, and migration costs while improving UX. The reward function is defined as follows.
\begin{equation}
    r:=\max_{L,E,C,\mathcal{E},\mathcal{U}}\mathbb{O}(t). 
\end{equation} For this multi-objective reward framework, the proposed MO-MADDPG algorithm effectively manages trade-offs and optimizes resource allocation and VT migration in real-time, enhancing both efficiency and user satisfaction within the dynamic multi-tier vehicular Metaverse, as presented below.

\subsubsection{Multi-Objective MADDPG Algorithm}
To handle the complexity and dynamics of the multi-objective optimization problem in the considered scenario, we adopt a MO-MADDPG. The MO-MADDPG algorithm is the extension of the MADDPG framework\cite{9903837}, which applies a policy gradient method for continuous action spaces that combines Q-learning with policy gradients, allowing for deterministic policy updates. In MO-MADDPG, agents have multiple objectives that they need to optimize simultaneously instead of a single objective. These objectives can sometimes be conflicting, requiring a trade-off. We utilize a \textit{Multi-Objective Actor-Critic (MOAC)} framework \cite{10113610} integrated with MADDPG, where agents optimize multiple objectives simultaneously while coordinating with others in a multi-agent environment. In this architecture, the actor serves as the policy network, which selects actions for the agent based on its current state. Meanwhile, the critic assesses these actions by estimating a value function that reflects the expected cumulative rewards, incorporating multiple objectives. To address the multi-objective nature, the reward function is expanded to include multiple goals, which can be either complementary or conflicting.  We apply the weighted sums technique to combine these objectives into a single scalar for optimization, with the agent's policies focusing on balancing the trade-offs between the various goals. Like MADDPG, MO-MADDPG follows the centralized training, decentralized execution framework, where each agent learns from the collective actions of all agents but acts independently during execution. In a multi-objective context, the key modifications involve the critic loss, which must evaluate multiple objectives at once; the actor updates, which need to manage the gradients from each objective's critic or prioritize trade-offs to optimize policy performance \cite{9498647}. This approach is particularly useful in environments where agents must collaborate while managing competing objectives, such as in resource allocation, autonomous driving, or energy management. The policy update in a MOAC involves taking the gradient of the combined reward with respect to the actor's parameters $\theta$. Let $\iota:=\{\iota_1,\iota_2,\iota_3,\iota_4,\iota_5\}$ is the set of multiple objectives, where $\iota_1,\iota_2,\iota_3$, $\iota_4$, and $\iota_5$ are UX, latency, EC, utility, and migration costs objectives, respectively. Thus,  in a weighted sum approach, the combined reward can be expressed as: $\mathcal{R}:=\sum_{\iota=1}^5\omega_{\iota}R_{\iota}$, where $R_{\iota}$ is the reward of $\iota$-th objective, and $\omega_{\iota}$ is the corresponding weight. The actor update is then given by:
\begin{equation}\label{actor}
    \nabla_{\theta}J(\theta):=\mathbb{E}_{\tau}\Bigg[\omega_{\iota}\nabla_{\theta}\log\pi_{\theta}(a_t|s_t)Q_x(s_t,a_t)\Bigg],
\end{equation}
where $\pi_{\theta}(a_t|s_t)$ is the policy function, $Q_{\iota}(s_t,a_t)$ is the $\iota$-th Q-value for the given objective at state $s_t$ and action $a_t$, $\tau$ represents the trajectory of state-action pairs, and $\mathbb{E}[\cdot]$ denotes expectation. On the other hand, for the critic network, there are multiple value functions, each corresponding to a different objective\cite{9903837}. The update rule is based on the difference between the target and current $Q$ values. For each objective $\iota$, the critic loss is updated as:
\begin{equation}
    \mathcal{L}_{\iota}(\phi_{\iota}):=\mathbb{E}_{s_t,a_t}\Bigg[\Big(Q_{\iota}\big(s_t,a_t\big)-\hat{Q}_{\iota}\big(s_t,a_t\big)\Big)^2\Bigg],
\end{equation}
where $\hat{Q}_{\iota}(s_t,a_t)$ is the target $Q$ value for the $\iota$-th objective, which can be calculated using the Bellman equation as:
\begin{equation}\label{bell}
    \hat{Q}_{\iota}(s_t,a_t):=r_t+\gamma\mathbb{E}_{s_{t+1}}\bigg[Q_{\iota}\big(s_{t+1},a_{t+1}\big)\bigg],
\end{equation}
where $r_t$ is the reward for the $\iota$-th objective, $\gamma$ is the discount factor, and $s_{t+1},a_{t+1}$ are the next state and action.
Therefore, the total critic loss is the sum of the losses from each objective\cite{9428174}, is expressed as:
\begin{equation}\label{criticloss}
    \mathcal{L}(\phi):=\frac{1}{5}\sum_{\iota=1}^5\mathcal{L}_{\iota}(\phi_{\iota}).
\end{equation}
The actor and critic are updated by computing the gradients of their respective objective losses:
$\nabla_{\theta}\mathcal{L}:=\frac{1}{5}\sum_{\iota=1}^5\nabla_{\theta}\omega_{\iota}Q_{\iota}(s_t,a_t)$.
The multi-objective framework modifies the standard actor-critic setup by introducing multiple value functions\cite{8616034}, each corresponding to a different objective, and adjusts the policy gradient using a weighted sum or trade-offs between objectives. These combined updates allow the actor and critic to manage and optimize multiple conflicting or complementary objectives effectively. 
The detailed procedure of MO-MADDPG scheme is given in \textbf{Algorithm \ref{algo1}}.\newline

\begin{algorithm}
\caption{MO-MADDPG Algorithm}
\label{algo1}
\begin{algorithmic}[1]
\algsetup{linenosize=\normalsize}
\footnotesize
\STATE Initialize actor and critic networks for each agent with weight parameters \(\theta_x\) and \(\phi_x\), respectively
\STATE Initialize target networks for actors and critics with weights \(\theta'_x\) and \(\phi'_x\)
\STATE Initialize replay buffer \(D\)
\STATE Initialize GCN parameters for spatial-temporal feature extraction
\STATE Initialize Stackelberg game parameters for decision-making
\STATE Initialize twin migration parameters
\FOR {each episode}
    \STATE Reset environment and obtain initial global state \(s_0\)
    \FOR {each time step \(t\)}
        \FOR {each agent \(x\)}
            \STATE Extract spatial and temporal features as GCN: \(z_x := \text{GCN}(s_x)\)
            \STATE Select action \(a_x\) using policy \(\pi_x(z_x)\) (with exploration)
        \ENDFOR
        \STATE Execute joint actions \(A := \big[a_1, a_2, \dots, a_X\big]\) in the environment
        \STATE Observe next state \(s'\), reward \(r\), and system metrics for each agent
        \STATE Store experience \((s, A, r, s')\) in replay buffer \(D\)
    \ENDFOR
    \FOR {each agent \(x\), sample a mini-batch from replay buffer \(D\)}
        \STATE Compute target Q-value using Bellman equation as in (\ref{bell})
        
        \STATE Update critic network using (\ref{criticloss})
      
    \ENDFOR
    \STATE Update actor network for each agent using (\ref{actor})
     \FOR {each agent, if migration criteria are met}
        \STATE Execute migration strategy using twin migration parameters
        \STATE Update Stackelberg game parameters for resource pricing and allocation as in (\ref{stg21}),(\ref{stg22}),(\ref{stg11}),(\ref{stg12})
    \ENDFOR
    \STATE Update GCN parameters using (\ref{eq-gcn})
    \STATE Refine multi-objective optimization parameters
\ENDFOR
\end{algorithmic}
\end{algorithm}

\section{Performance Evaluation}
The proposed MO-MADDPG scheme is compared with state-of-the-art approaches such as the MADQN\cite{10742925}, MADDPG\cite{10521838}, and a genetic algorithm (GA)\cite{10185562}. 

We conducted simulations in a Python 3.6 environment on a computer with a 2.4 GHz Core i7 CPU and 16GB RAM. The simulation models a three-tier network consisting of 100 vehicles, 10 edge servers, and 3 cloud servers. Vehicles are assumed to move at speeds between 36 and 108 km/hr and their positions update every second. VT task size is [10, 50]MB, requiring CPU cycles in [0.6, 1.6] Gcycles. Edge and cloud servers' computing power is in the range of [10, 50] W and [50, 200] W, respectively. Migration cost is in the range of [0.30, 0.50] \$ per MB of the transferred VT tasks, edge and cloud servers' allocated bandwidth are 20 MHz and 100 MHz, respectively. The learning rates for the critic and actor networks are \(10^{-3}\) and \(10^{-4}\), respectively. The number of episodes is 3000, with 100 steps per episode, a discount factor of 0.95, a replay memory size of \(10^5\), a mini-batch size of 128, and a soft update factor of \(1 \times 10^{-3}\). The system channel bandwidth is set to 60 MHz.

\begin{figure}[t]
 \centering
       \begin{subfigure}{0.23\textwidth}
    \includegraphics[width=\textwidth]{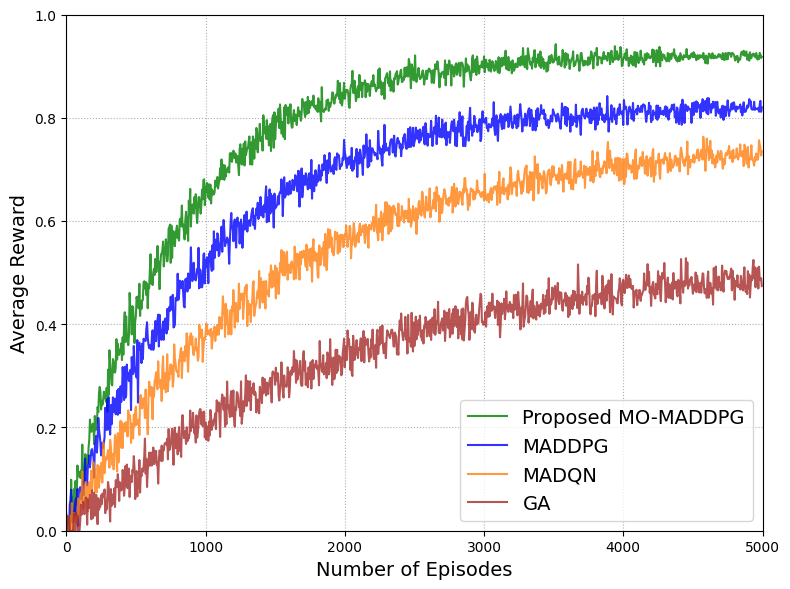}
         \caption{}
        \label{rewad}
        \end{subfigure}%
        \begin{subfigure}{0.23\textwidth}
    \includegraphics[width=\textwidth]{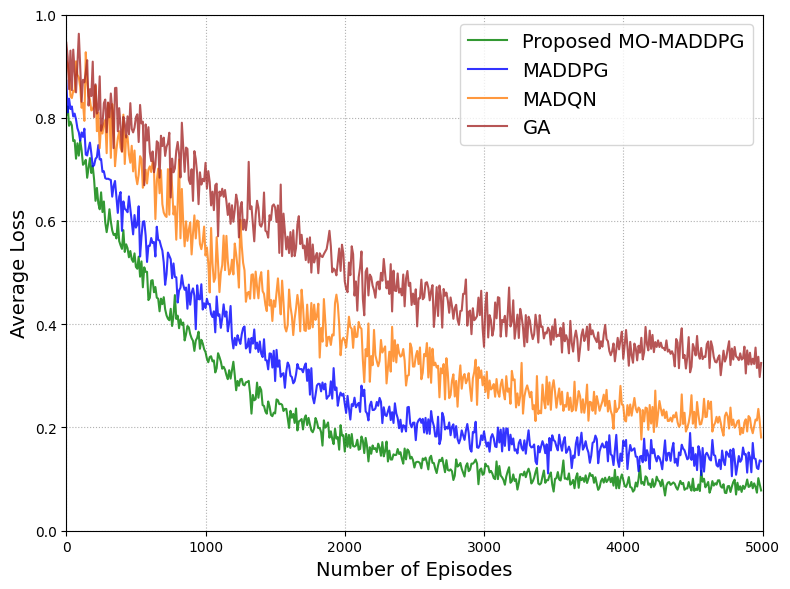}
         \caption{}
        \label{loss}
        \end{subfigure}%
\caption{Convergence analysis of the considered algorithms: a) Reward vs episodes; b) Loss value vs episodes.}
    \label{fig:conv}
\end{figure}

\emph{1) Convergency Evaluation:} We assessed the convergence performance of all compared algorithms in terms of reward and loss value. As presented in Fig. \ref{rewad}, MO-MADDPG demonstrates a significantly faster and higher convergence in terms of reward, reflecting its ability to optimize multiple conflicting objectives, such as latency, EC, migration costs, and UX. The MO-MADDPG’s advanced decision-making strategies ensure a balanced trade-off among these objectives, resulting in a more efficient resource allocation and VT migration process. MADDPG, while competitive, lacks the advanced multi-objective design of MO-MADDPG, limiting its adaptability to conflicting objectives. MADQN demonstrates slower convergence and lower rewards due to its reliance on discrete decision-making, which is less effective in dynamic environments with continuous action-state spaces. As a non-learning-based approach, GA demonstrates the lowest performance, marked by slow convergence, owing to its lack of adaptability to real-time vehicular dynamics. 

In terms of loss value minimization, our proposed MO-MADDPG outperforms other methods by achieving faster and more stable convergence, as shown in Fig. \ref{loss}. It effectively reduces loss value through its hierarchical and multi-agent framework, underscoring its adaptability and efficiency in dynamic vehicular Metaverse environments. MADDPG achieves moderate loss reduction but is limited by its single-objective optimization approach. MADQN struggles to reduce loss values due to its discrete framework, while GA performs poorly as it relies on static optimization rather than adaptive learning. Overall, MO-MADDPG outperforms other methods, achieving up to 32\% higher rewards and 25\% lower loss values compared to alternative approaches. This performance is attributed to the integration of GCNs and a hierarchical Stackelberg game incentive scheme, which enhances its scalability and adaptability, making it highly effective in multi-tier vehicular environments. In comparison, the other approaches fall short due to the lack of such useful integration of frameworks, limiting their performance.
 
\begin{figure}[t]
 \centering
       \begin{subfigure}{0.23\textwidth}
    \includegraphics[width=\textwidth]{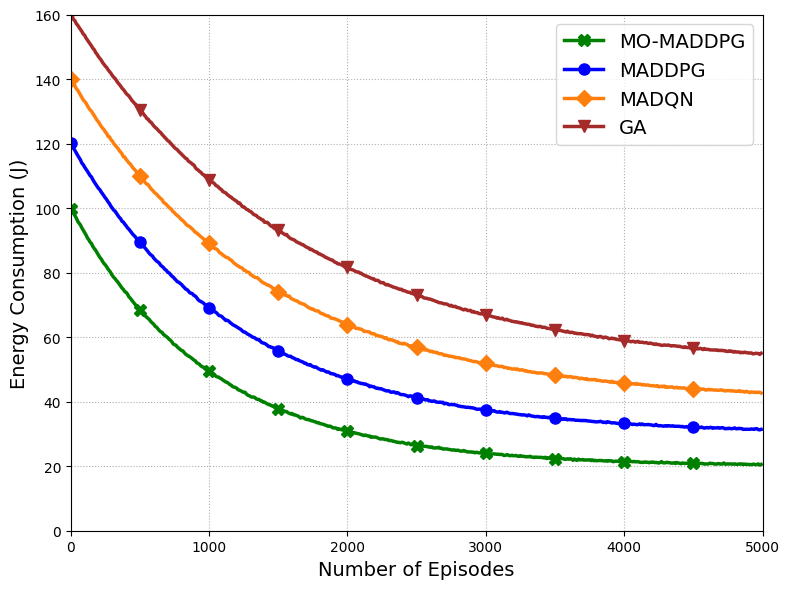}
         \caption{}
        \label{EC1}
        \end{subfigure}%
        \begin{subfigure}{0.23\textwidth}
    \includegraphics[width=\textwidth]{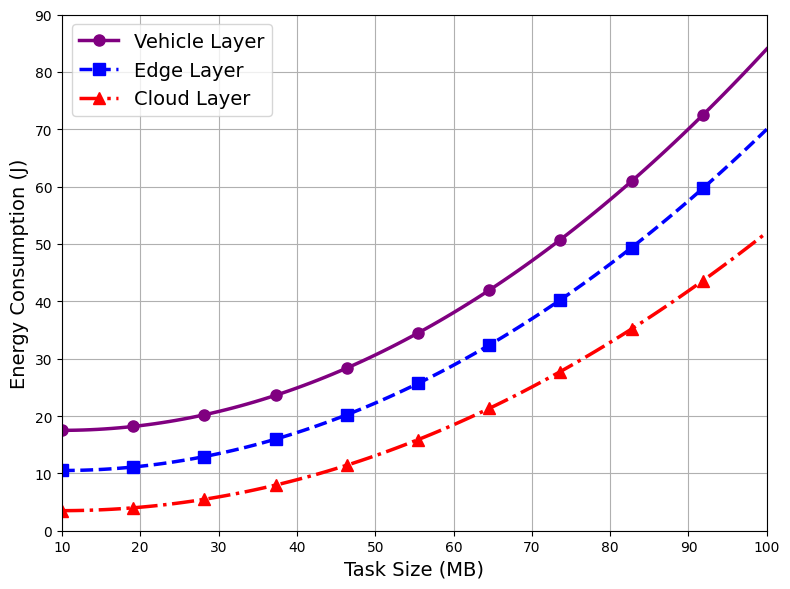}
         \caption{}
        \label{EC2}
        \end{subfigure}%
\caption{Analysis of EC across layers for different algorithms: a) EC vs episodes; b) EC vs task size.}
    \label{EC}
\end{figure}

\emph{2) EC:} We evaluate the EC of all compared algorithms, including the proposed MO-MADDPG, MADQN, MADDPG, and the non-learning GA across different layers (vehicle, edge, and cloud),  as shown in Fig. \ref{EC}. As the number of episodes increases, EC decreases for all algorithms, indicating improved efficiency in resource allocation and VT migration, as shown in Fig. \ref{EC1}. This trend indicates that as the algorithms train and adapt over time, they become more efficient in managing resource allocation and VT migration. Among the algorithms, the proposed MO-MADDPG achieves the lowest EC, showcasing better optimization capabilities. While other algorithms -- such as MADQN and MADDPG -- also exhibit reduced EC over time, they remain less efficient than MO-MADDPG. The GA approach shows minimal improvement, underscoring its limitations in handling complex and dynamic environments. This overall reduction in EC underscores the importance of learning-based approaches in enhancing system efficiency. 

We also analyze how EC changes across various layers as task sizes increase for the proposed algorithm. As shown in Fig. \ref{EC2}, this evaluation reveals that EC rises across all layers with larger task sizes, owing to the greater computational and communication demands. As task size increases, energy consumption varies significantly across layers. In the vehicle layer, energy consumption increases significantly, reaching up to 50\% for large tasks. This is primarily due to the limited computational capacity of vehicles and their reliance on battery power. In the edge layer, energy consumption grows more moderately, peaking at 26\% as task size increases. This is because edge servers efficiently balance computation and proximity. In the cloud layer, processing energy remains lower and stable, but transmission energy can dominate, resulting in a modest overall increase of 17\% for large tasks.

\begin{figure}[t]
 \centering
       \begin{subfigure}{0.23\textwidth}
    \includegraphics[width=\textwidth]{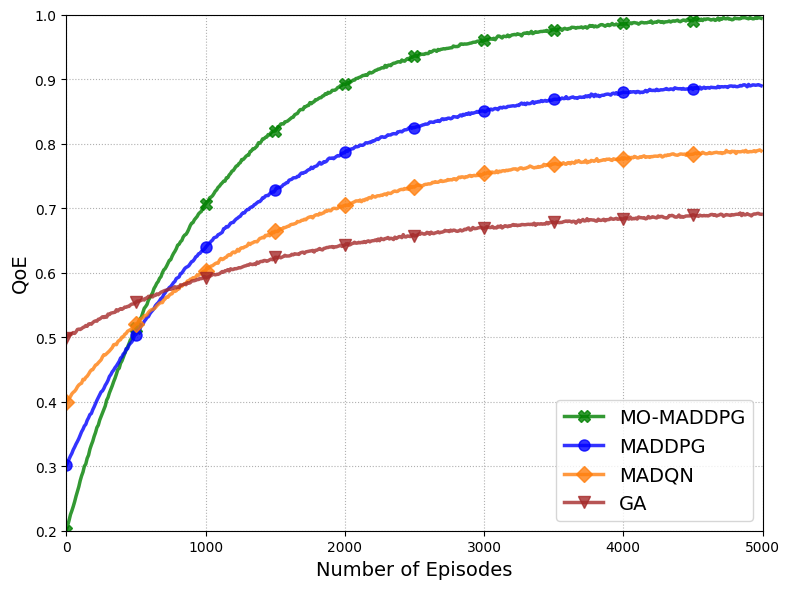}
         \caption{}
        \label{QoE1}
        \end{subfigure}%
        \begin{subfigure}{0.23\textwidth}
    \includegraphics[width=\textwidth]{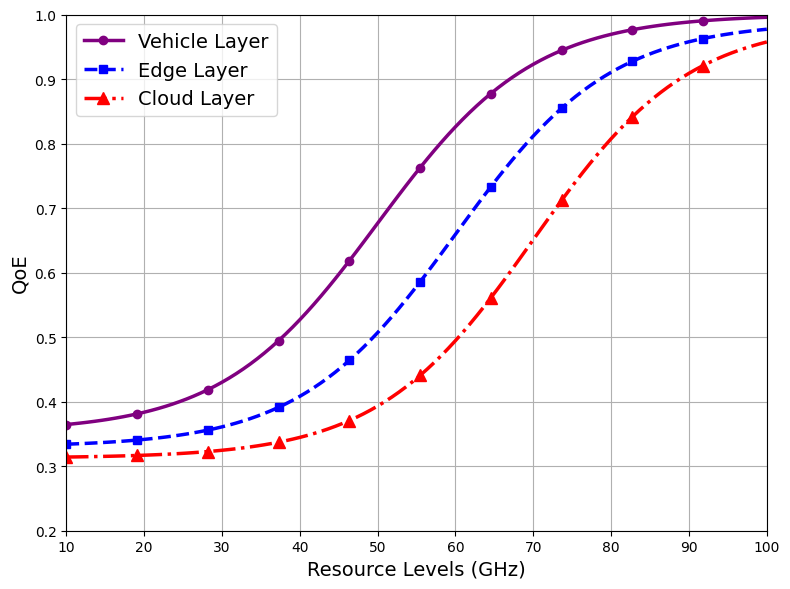}
         \caption{}
        \label{QoE2}
        \end{subfigure}%
\caption{QoE Comparison across different layers: a) QoE vs episode; b) QoE vs available resources.}
    \label{fig:QoE}
\end{figure}

\emph{3) Quality of Experience:}
We assess the QoE performance of the compared algorithms, including MO-MADDPG, MADDPG, MADQN, and GA, as the number of learning episodes increases, as illustrated in Fig. \ref{QoE1}. The proposed MO-MADDPG achieves the highest QoE than the MADDPG, MADQN, and GA schemes. Because MO-MADDPG is designed to balance multiple conflicting objectives such as latency, EC, migration costs, and user satisfaction. This multi-objective optimization capability is achieved with the integration of GCNs and hierarchical Stackelberg incentives, allowing it to adapt dynamically to changing conditions.
In comparison, MADDPG performs moderately well, showing improvements over MADQN and GA but falling short of MO-MADDPG due to its lack of multi-objective capabilities. MADQN, reliant on discrete decision-making, exhibits slower QoE improvement and struggles in dynamic and continuous environments. The GA algorithm, being non-adaptive, achieves the lowest QoE, highlighting its inability to respond effectively to real-time changes.

We also evaluate the QoE performance of all algorithms as resource availability at each layer increases, as shown in Fig. \ref{QoE2}. All algorithms demonstrate improved QoE. However, MO-MADDPG excels in utilizing resources efficiently, ensuring optimal performance even under constrained conditions, unlike GA and MADQN, which fail to fully capitalize on the available resources. This analysis shows that MO-MADDPG provides a robust and flexible framework for optimizing QoE in the multi-tier vehicular Metaverse.

\begin{figure*}[t]
 \centering
       \begin{subfigure}{0.24\textwidth}
    \includegraphics[width=\textwidth]{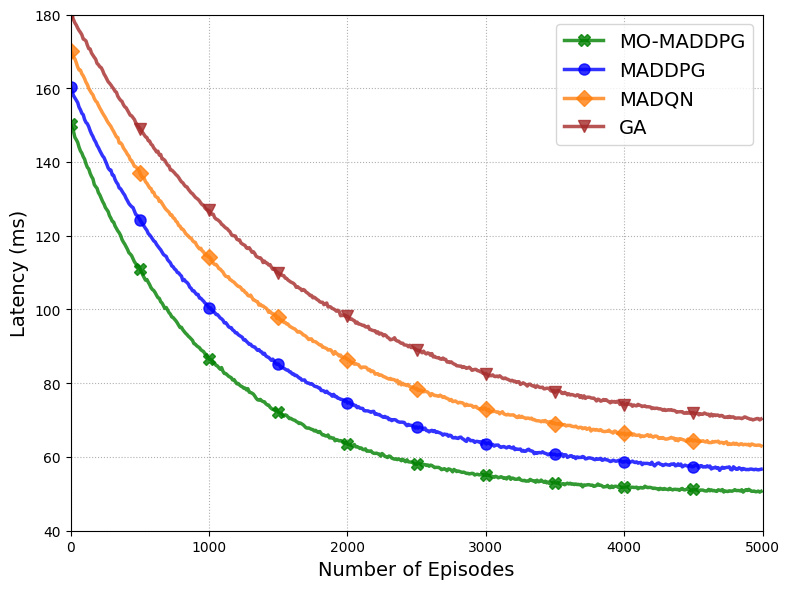}
         \caption{}
        \label{latency1}
        \end{subfigure}%
        \begin{subfigure}{0.24\textwidth}
    \includegraphics[width=\textwidth]{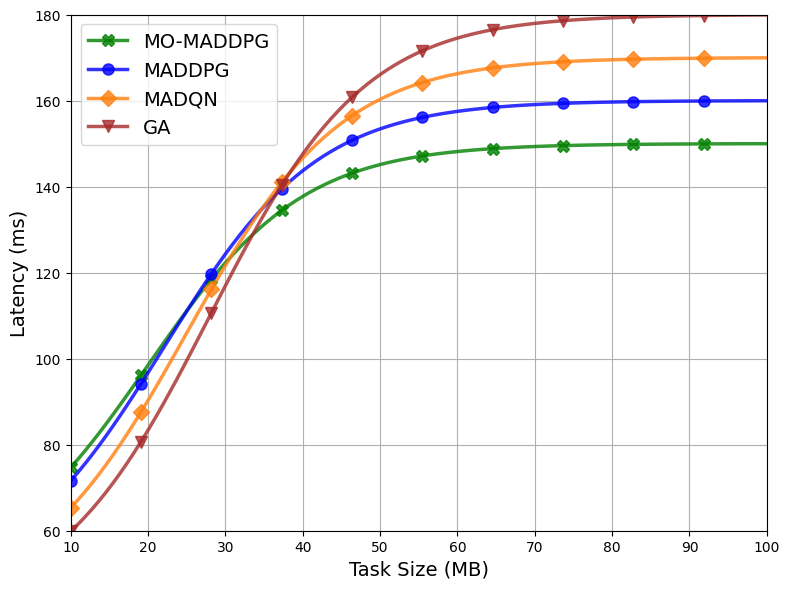}
         \caption{}
        \label{latency2}
        \end{subfigure}%
        \begin{subfigure}{0.24\textwidth}
    \includegraphics[width=\textwidth]{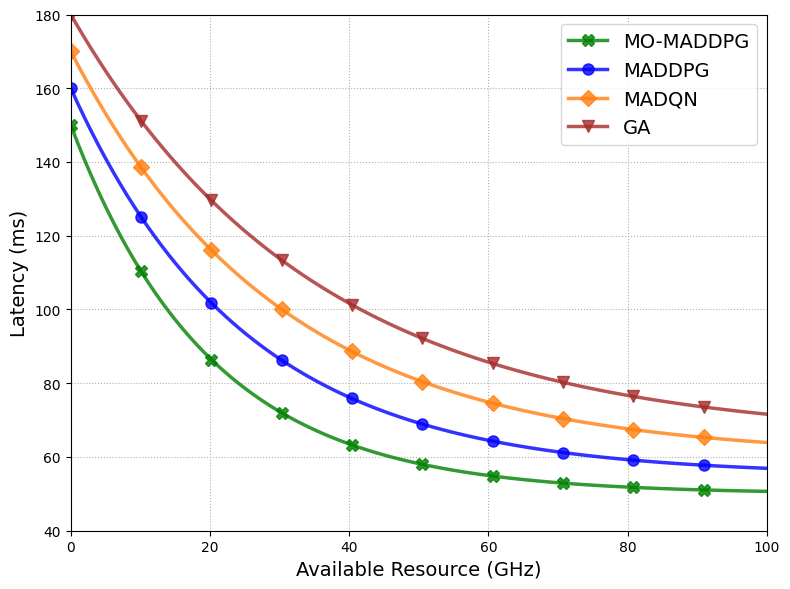}
         \caption{}
        \label{latency3}
        \end{subfigure}%
         \begin{subfigure}{0.24\textwidth}
    \includegraphics[width=\textwidth]{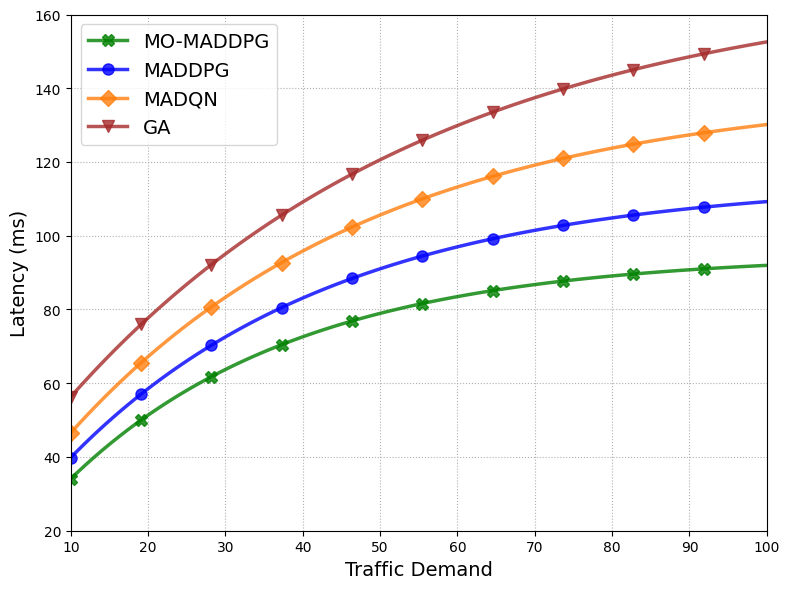}
         \caption{}
        \label{latency4}
        \end{subfigure}%
\caption{Latency comparison across algorithms at different layers: a) Latency vs episode; b) Latency vs task size; c) Latency vs resource level; d) Latency vs traffic demand.}
    \label{fig:latency}
\end{figure*}

\emph{4) Latency Evaluation:} We examined the VT task migration latency across all compared algorithms considering learning episodes, task size, resource availability, and VT task demand. Fig.\ref{latency1} illustrates a reduction in latency with an increase in learning episodes. The results indicate that MO-MADDPG excels in rapidly reducing latency with increased training episodes. In contrast, MADDPG and MADQN show slower latency reductions, pointing to less efficient optimization mechanisms, while GA provides the highest latency values throughout the episodes. This indicates MO-MADDPG's strong learning capabilities, enabling it to adapt effectively to dynamic environments while empowering agents at each layer to make intelligent decisions on resource allocation and VT migration. Similarly, we evaluate the VT migration latency with respect to increasing VT task size as shown in Fig. \ref{latency2}. As task size increases, all algorithms experience higher latency due to greater computational and communication demands. However, MO-MADDPG consistently maintains the lowest latency across different task sizes, demonstrating its efficiency in managing larger workloads. On the other hand, MADDPG and MADQN struggle to maintain consistent performance, while GA demonstrates significant inefficiency, with latency increasing sharply as task sizes increase.

Fig. \ref{latency3} illustrates the VT migration latency trend as resources available at each layer of the hierarchy increase increases. The MO-MADDPG algorithm ensures efficient resource allocation and VT migration decisions, outperforming other methods with its ability to achieve minimal latency. MADDPG and MADQN show moderate reductions in latency with increasing resource availability but are still outperformed by MO-MADDPG. In contrast, GA remains the least efficient, exhibiting the highest latency among all algorithms even as resources increase.

Furthermore, we evaluated the VT migration latency under increasing migration and resource allocation demand. As shown in Fig. \ref{latency4}, the higher VT migration demand results in increased latency across all algorithms. However, MO-MADDPG handles this growth more efficiently, maintaining significantly lower latency even in high-demand scenarios. This signifies its robustness in adapting to complex network conditions. While MADDPG and MADQN perform reasonably well, their latency rises more sharply as demand grows. GA, being static and non-adaptive, performs the worst, with latency surging dramatically as demand increases.
 Overall, MO-MADDPG excels in optimizing latency across various conditions, thanks to its advanced multi-objective optimization, GCN integration, and game-theoretic incentives, consistently outperforming traditional and non-learning methods.

\begin{figure*}[t]
 \centering
       \begin{subfigure}{0.28\textwidth}
    \includegraphics[width=\textwidth]{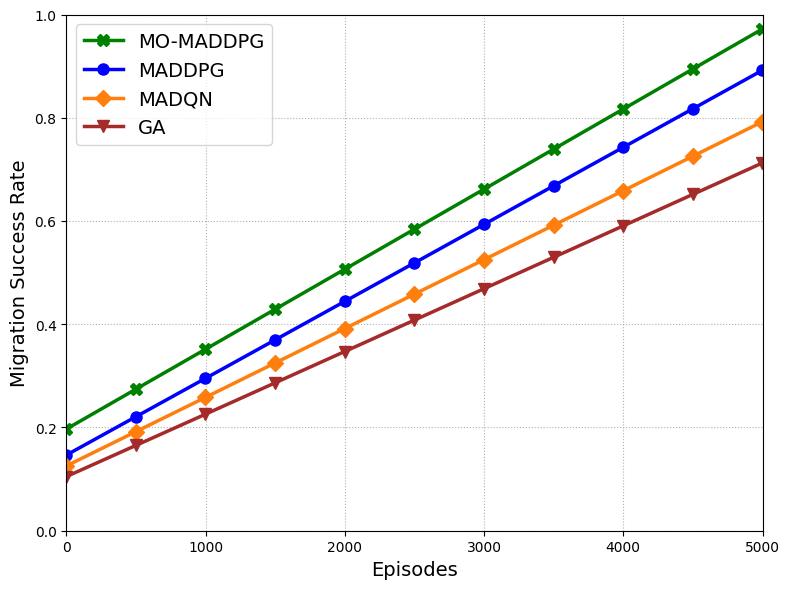}
         \caption{}
        \label{MSR1}
        \end{subfigure}%
        \begin{subfigure}{0.28\textwidth}
    \includegraphics[width=\textwidth]{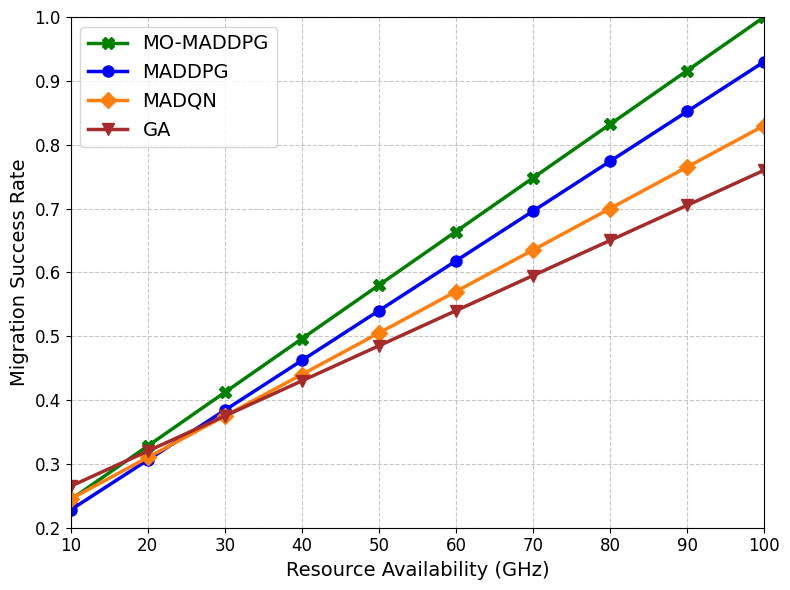}
         \caption{}
        \label{MSR2}
        \end{subfigure}%
        \begin{subfigure}{0.28\textwidth}
    \includegraphics[width=\textwidth]{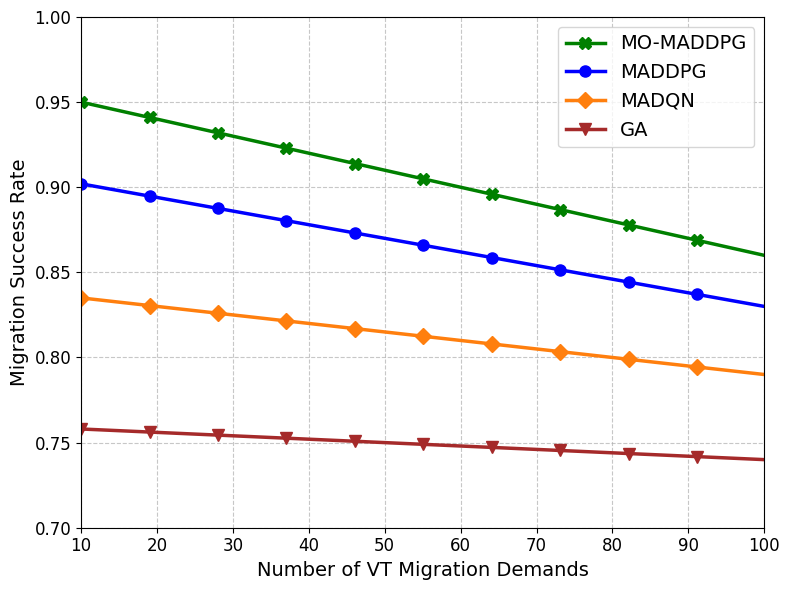}
         \caption{}
        \label{MSR3}
        \end{subfigure}%
    \caption{Migration success rate comparison across on different layers: a) MSR vs episode; b) MSR vs available resource level; c) MSR vs VT migration demand.}
    \label{MSR}
\end{figure*}

\emph{5) Migration Success Rate:}
We assess the migration success rates of MO-MADDPG, MADDPG, MADQN, and GA under varying conditions, including increasing episodes, resource availability, and VT migration demand. As the number of episodes increases, the migration success rate increases for all compared algorithms, as shown in Fig. \ref{MSR1}. However, MO-MADDPG achieves a steady and significant improvement in migration success rates due to its integration of advanced multi-objective optimization techniques, GCNs, and a hierarchical Stackelberg game incentive scheme. In contrast, MADDPG and MADQN are less effective due to the lack of these advanced features and their reliance on single-objective optimization. Furthermore, MADQN's use of discrete decision-making frameworks restricts its ability to adapt to dynamic conditions, handle complex resource dependencies, and ensure consistent migration success. The GA algorithm, on the other hand, demonstrates the lowest migration success rates, as it lacks adaptability and relies on static optimization.
As resource availability increases at each layer of the hierarchy, MO-MADDPG consistently achieves higher migration success rates compared to other methods as shown in Fig. \ref{MSR2}. This is because the integration of GCNs in its framework allows for accurate predictions of resource demands and optimal migration paths, ensuring optimal VT migration and resource utilization decisions. In comparison, MADDPG and MADQN show reduced performance as resource availability increases, while GA struggles to adapt to the added complexity, leading to poor success rates.
 Moreover, under increasing migration demand, MO-MADDPG maintains high migration success rates, as illustrated in Fig. \ref{MSR3}. This highlights its scalability and resilience in dynamic, high-demand scenarios, achieved through its hierarchical Stackelberg game-based incentive mechanism. This approach aligns the objectives of vehicles, edge nodes, and cloud servers, facilitating efficient resource allocation and migration decisions. In contrast, MADDPG and MADQN exhibit moderate declines in success rates as demand increases, while GA's static optimization method results in significant underperformance. In summary, MO-MADDPG's advanced learning techniques, multi-objective optimization, GCNs, and game-theoretic incentives enable it to achieve up to 30-35\% higher migration success rates compared to other methods, excelling under diverse and challenging vehicular Metaverse conditions.

\begin{figure}[!ht]
 \centering
       \begin{subfigure}{0.23\textwidth}
    \includegraphics[width=\textwidth]{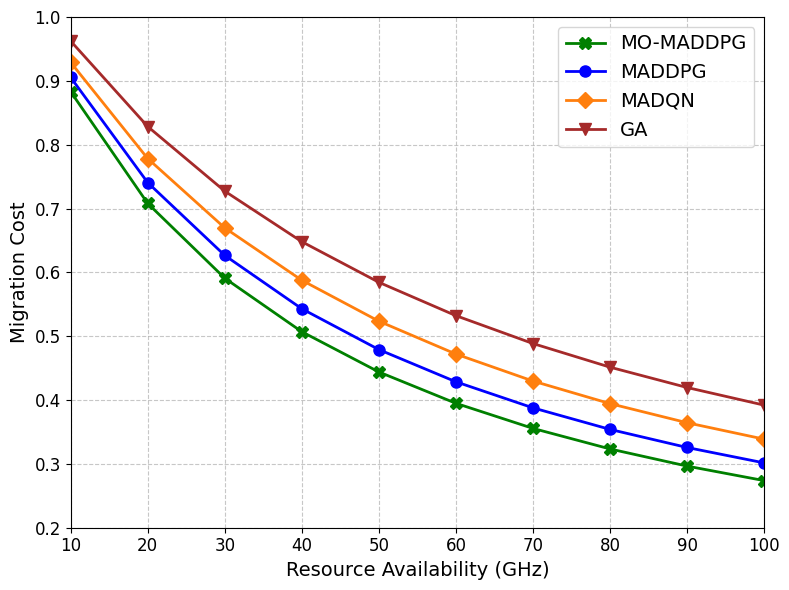}
         \caption{}
        \label{cost1}
        \end{subfigure}%
        \begin{subfigure}{0.23\textwidth}
    \includegraphics[width=\textwidth]{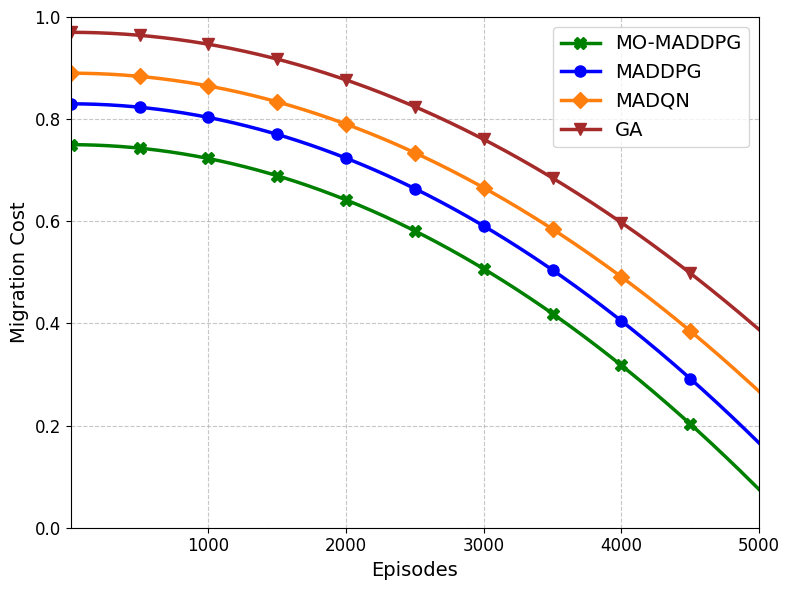}
         \caption{}
        \label{cost2}
        \end{subfigure}%
\caption{Migration cost comparison on different layers: a) Cost vs resource level; b) Cost vs episode.}
    \label{fig:cost}
\end{figure}

\emph{6) Migration Cost Evaluation:} 
We examine and compare the migration costs of various algorithms, including MO-MADDPG, MADDPG, MADQN, and GA, under different resource levels and learning episodes. As resource levels increase, migration costs generally decrease for all algorithms, as shown in Fig. \ref{cost1}, reflecting improved efficiency in resource allocation and seamless VT migration. The results demonstrate that MO-MADDPG achieves significantly lower migration costs compared to other algorithms. This is attributed to its advanced multi-objective optimization capabilities, which utilize GCNs and hierarchical Stackelberg game-based incentive mechanisms to dynamically allocate resources and optimize VT migration paths. 

As learning episodes increase, all algorithms demonstrate a reduction in migration costs, but MO-MADDPG exhibits the fastest convergence and lowest final costs. This is due to its ability to balance conflicting objectives, such as latency, EC, and UX, while simultaneously minimizing migration costs. In contrast, MADDPG and MADQN show slower convergence and higher costs, as it lacks advanced multi-objective optimization features. GA, being a static optimization method, incurs significantly higher costs and faces challenges in adapting to the dynamic nature of vehicular networks. Overall, MO-MADDPG achieves up to 30\% lower migration costs than other algorithms by leveraging GCNs for spatial-temporal insights, MADRL-based strategies for trade-off optimization, and hierarchical decision-making for efficient resource use in the resource-constrained vehicular Metaverse.

\section{Concluding Remarks}
The Metaverse finds a myriad of applications in education, smart city and smart home, entertainment, modeling and monotoring, autonomous driving, medicine, tourism, business, real estate, socialization, and manufacturing. Among the various types of Metaverses, the vehicular Metaverse promises to provide immersive vehicular services and experiences by seamlessly integrating the Metaverse into vehicular networks. Because the vehicular Metaverse should effectively enable each VMU to seamlessly connect with its VT, optimal resource allocation and VT migration are important research problems worth investigating. Concerning optimal resource allocation and VT migration in the vehicular Metaverse, we proposed a novel multi-tier resource allocation and VT migration framework to address the complex challenges of VT migration and resource management in the multi-tier vehicular Metaverse. The proposed framework unifies GCN, a hierarchical Stackelberg game-based incentive mechanism, and MADRL to enhance resource utilization, reduce latency, improve migration success, and optimize user experience. The GCN captures spatial-temporal dependencies in vehicular networks and the MADRL-based algorithm, MO-MADDPG, integrates global and local state information to facilitate informed decision-making across vehicles, edge nodes, and cloud servers, enabling a scalable and adaptive system. The developed approach enables vehicles to autonomously offload and migrate VT tasks; dynamically allocate resources; and balance loads across vehicle, edge, and cloud layers, while alleviating computational bottlenecks at the edge. We adopt a Stackelberg game-based incentive mechanism to foster collaboration among entities at all tiers, ensuring efficient and seamless operation in a dynamic, resource-constrained environment. We conducted extensive simulations to validate the effectiveness of the proposed framework and demonstrated its ability in optimizing system performance under varying network conditions and mobility scenarios while enhancing the reliability and scalability of vehicular Metaverse services to meet stringent QoS requirements.

\bibliographystyle{IEEEtran}
\bibliography{IEEEabrv,References}

\end{document}